\documentclass[journal]{IEEEtran}
\usepackage{amsfonts}
\usepackage{makecell}
\usepackage{graphicx}        
\usepackage{float} 
\usepackage{pifont}
\usepackage{caption}
\usepackage{multirow}
\usepackage[caption=false,font=normalsize,labelfont=sf,textfont=sf]{subfig}
\usepackage[table,xcdraw]{xcolor}

\usepackage{cite}

\ifCLASSINFOpdf
\else
\fi
\usepackage{amsmath,amsfonts,amsthm,bm}
\usepackage{array}
\usepackage{url}

\begin{document}
%
\title{VolumeNet: A Lightweight Parallel Network for Super-Resolution of Medical Volumetric Data}
%
%
%

\author{Yinhao~Li,
		Yutaro~Iwamoto,
		Lanfen~Lin,
		Rui~Xu,
    and~Yen-Wei~Chen,~\IEEEmembership{Member,~IEEE}
\thanks{Yinhao Li is with the Graduate School of Information Science and Engineering, Ritsumeikan University, Kusatsu, Japan (e-mail: gr0278ps@ed.ritsumei.ac.jp).}
\thanks{Yutaro Iwamoto is with the College of Information Science and Engineering, Ritsumeikan University, Kusatsu, Japan (e-mail: yiwamoto@fc.ritsumei.ac.jp).}
\thanks{Lanfen Lin is with the  College of Computer Science, Zhejiang University, China  (e-mail: llf@zju.edu.cn).}
\thanks{Rui Xu is with the DUT-RU international school of information science and engineering, DUT-RU Co-Research Center of Advanced ICT for Active Life, Key Laboratory for Ubiquitous Network and Service Software of Liaoning Province, Dalian university of technology, China (e-mail: xurui@dlut.edu.cn).}
\thanks{Yen-Wei Chen is with the College of Information Science and Engineering, Ritsumeikan University, Japan, the College of Computer Science and Technology, Zhejiang University, China, and the Research Center for Healthcare Data Science, Zhejiang Lab, China (e-mail: chen@is.ritsumei.ac.jp).}
\thanks{Corresponding authors: Yen-Wei Chen and Lanfen Lin}
}

\maketitle

\begin{abstract}
Deep learning-based super-resolution (SR) techniques have generally achieved excellent performance in the computer vision field. Recently, it has been proven that three-dimensional (3D) SR for medical volumetric data delivers better visual results than conventional two-dimensional (2D) processing. However, deepening and widening 3D networks increases training difficulty significantly due to the large number of parameters and small number of training samples. Thus, we propose a 3D convolutional neural network (CNN) for SR of medical volumetric data called ParallelNet using parallel connections. We construct a parallel connection structure based on the group convolution and feature aggregation to build a 3D CNN that is as wide as possible with a few parameters. As a result, the model thoroughly learns more feature maps with larger receptive fields. In addition, to further improve accuracy, we present an efficient version of ParallelNet (called VolumeNet), which reduces the number of parameters and deepens ParallelNet using a proposed lightweight building block module called the Queue module. Unlike most lightweight CNNs based on depthwise convolutions, the Queue module is primarily constructed using separable 2D cross-channel convolutions. As a result, the number of network parameters and computational complexity can be reduced significantly while maintaining accuracy due to full channel fusion. Experimental results demonstrate that the proposed VolumeNet significantly reduces the number of model parameters and achieves high precision results compared to state-of-the-art methods.
\end{abstract}


%
\IEEEpeerreviewmaketitle

\section{Introduction}
%
%
%
%
\IEEEPARstart{H}{igh-resolution} 
(HR) images captured via magnetic resonance imaging (MRI) or computerized tomography (CT) are generally preferable in clinical diagnosis because they provide rich texture details and enhance the fidelity of radiomic features, which promotes accurate subsequent analysis and early diagnosis. However, quickly obtaining precise medical data with a higher spatial resolution is a challenging task as spatial resolution increases, the signal-to-noise ratio is reduced, and scan time increases, which makes it impracticable for most MR equipment in hospitals to operate in clinical environments. One solution for this issue is to use SR technique. Especially, the single image SR (SISR) only requires a low-resolution (LR) scan to restore HR output without extra scan time, which may overcome the identified drawbacks. To date, SISR has been used for accurate medical image diagnosis and analysis tasks such as brain tumor detection \cite{ozyurt2020expert}, dental applications \cite{hatvani2018tensor}, retinal blood vessel segmentation \cite{mahapatra2019image}, and retinal disease detection \cite{anoop2020super}. There are also many researches conducting on development of medical image SR techniques including 2D and 3D SR techniques \cite{pham2017brain, chen2018brain, chen2018efficient, zhu2019can, zhao2019channel, you2019ct, du2020super, lyu2020mri, du2018accelerated}. The purpose of this work is to improve the accuracy and efficiency of 3D SISR for medical volume.

Deep learning-based SR methods, which are a practical natural image restoration approaches, have recently attracted increasing attention due to their superior performance \cite{dong2015image, kim2016accurate, shi2016real, zhang2018residual, zhang2018image, mao2016image, han2018image, li2018multi, huang2017densely, ledig2017photo, zhang2018learning, haris2018deep, tong2017image, kim2016deeply, lai2018fast, lai2017deep, dahl2017pixel, ren2017image, wang2018fully, bae2017beyond, guo2017deep, huang2017wavelet, liu2018multi, wang2020deep}. Deepening and widening the network was demonstrated to significantly improve convolutional neural network (CNN) accuracy significantly in \cite{tan2019efficientnet}, because deeper and wider models with complex structures can fully extract and merge hierarchical feature maps. However, differing from ordinary 2D natural images, medical images (i.e., 3D volume data) have continuous spatial information in all directions, and the quality of medical volume data depends on the quality of the volume reconstruction from a set of slice images. In contrast, 2D CNNs only work slice by slice, i.e., information from continuous structures in the third dimension is discarded. However, training and testing deep and wide 3D CNN models is challenging because such models have large numbers of parameters, consume more memory, and are more computationally expensive.

Due to the hardware limitation, most previous studies focused on solving in-plain or through-plain SR using 2D CNNs \cite{zhao2018self, shi2018mr, zhao2019channel, zhao2016deep, hu2018single, zhu2019can, you2019ct, du2020super, lyu2020mri}; however, some studies have attempted to accomplish both in-plain and through-plain SR simultaneously. For example, Pham et al., who first used a 3D CNN for MRI image SR, presented a 3D CNN with only three layers \cite{pham2017brain}. In addition, Chen et al. proposed densely connected SR network (DCSRN) \cite{chen2018brain} and multi-level DCSRN (mDCSRN) \cite{chen2018efficient} using 3D dense connections. They realized good performance with a few parameters. However, the network with a small number of layers is insufficient relative to obtaining satisfactory accuracy. In addition, models based on the generative adversarial network (GAN) tend to yield fake details, which is a critical problem in medical image processing. Simply constructing a shallow 3D CNN or directly converting 2D models to 3D models cannot improve the accuracy and efficiency of existing methods simultaneously. Thus, \textit{we propose an accurate end-to-end system for medical volumetric data SR, which we refer to as the 3D parallel-connected network (ParallelNet). In the proposed ParallelNet, hierarchical features are fully fused throughout the wider network structure}. 

Under the constraint of a limited number of training samples, it is difficult to increase the depth of a 3D CNN to improve accuracy and avoid overfitting. The method based on a model using prior knowledge \cite{cherukuri2019deep} is effective but increases computational complexity. Thus, we aim to propose a novel high-efficiency network with lightweight modules to reduce the overall computational costs of deep learning methods. Relative to using lightweight models to realize faster predictions, many studies have attempted to develop efficient deep learning models using separable convolutions to process 2D images immediately with low power consumption \cite{rigamonti2013learning, kim2015compression, lebedev2014speeding, howard2017mobilenets, chollet2017xception, wang2017factorized, jin2014flattened, tai2017memnet, tai2017image, hui2018fast, ahn2018fast}. In such lightweight models, depthwise and pointwise convolutions are commonly used rather than standard full convolution; however, most depthwise convolution-based methods for detection or classification tasks tend to lose accuracy due to insufficient integration between channels. Thus, they are not suitable for medical data SR. A tensor factorization method \cite{hatvani2018tensor} was proposed to enhance 3D image resolution by decomposing the image of interest using canonical polyadic decomposition (CPD). However, The CPD-based lightweight CNN model reduce accuracy due to extensive use of the depthwise convolution. Therefore, \textit{we also present an improved version of ParallelNet (called VolumeNet) that reduces the number of parameters and increases the depth of ParallelNet using a lightweight module, which enhances the capability of our model in terms of accuracy and efficiency}. The proposed methods contribute the following technical achievements:

\textbf{Wider and more accurate backbone network.} Inspired by group convolution, we propose the parallel connection structure. Here, by stacking group convolutions in parallel, the network can be widened significantly and the number of model parameters can avoid explosive growth. In addition, a feature aggregation module is presented to integrate the feature maps of each branch, which helps avoid loss of accuracy caused by a lack of information fusion between channels.

\textbf{Deeper and more efficient building block.} To further improve the accuracy and efficiency of the proposed ParallelNet, we propose a lightweight structure (the Queue module), which is inspired by Xception \cite{hui2018fast} and  MobileNetV2 \cite{sandler2018mobilenetv2}. The proposed Queue module comprises pointwise convolution for shrinking channels, three separable 2D convolutions for feature mapping, and pointwise convolution for expanding channels. We use the proposed Queue module rather than a standard convolution kernel for feature extraction and feature fusion. Note that, in the proposed lightweight Queue module, we employ cross-channel one-dimensional (1D) convolution (i.e., 2D convolution) rather than depthwise convolution. As a result, so the information of all feature maps in each layer can be fully integrated, and the model can be deepened without loss of accuracy. To use (2+1)D convolution layers instead of 3D convolution layers is also a good idea \cite{tran2018closer}, but using 1D convolution layers may reduce more parameters. In addition, we found in \cite{tran2018closer} that, in the feature maps of each layer, the resolution of each slice is much larger than the resolution of the slice direction, so this method may not perform well in our research. The shrinking-and-expansion structure in the Queue module allows most of the convolutional operations to be performed in low-dimensional space. Thus, the number of network parameters and computational complexity can be reduced significantly. In addition, the proposed lightweight Queue module are versatile and can be combined with a wide variety of existing network architectures and prior knowledge \cite{cherukuri2019deep} to further enhance performance. 

By converting the proposed methods to 2D, the methods can be applied to a combination of 2D SR on sinograms and volume reconstruction, to e.g., solve the sinogram SR problem reported in the literature \cite{awasthi2020deep}. Original 2D sinogram data are difficult to obtain, and artifacts are generated during volume reconstruction; thus, to the best of our knowledge, few studies have investigated SR for medical volumetric data based on sinogram data.
\\
\textbf{Contributions:} The goal of this study is to present a 3D SR CNN structure with lightweight building blocks to realize more efficient processing and accurate results; thus, our primary contributions are summarized as follows:

•	\textbf{3D CNN Structure for Volumetric Data SR:} We propose a deep 3D CNN backbone for medical volumetric data SR that has a deeper and wider structure compared to most state-of-the-art methods. In addition, the proposed model fully integrates hierarchical feature maps, particularly feature maps with large respective fields. Through the SR results of MRI and CT images, we demonstrate that the proposed model outperforms state-of-the-art methods in terms of accuracy. In addition, the experimental results prove that the proposed method improves the processing speed in the inference phase.

•	\textbf{Lightweight CNN Module for Fast and Accurate 3D SR:} We propose a lightweight module that can effectively speed up processing by reducing the number of parameters and improve accuracy by deepening the network. We demonstrate that the proposed method can also be combined with other arbitrary SR networks to improve accuracy.

•	\textbf{Method Comparison:} To demonstrate the effectiveness of the proposed method, we compare it to several 3D SR CNNs accelerated by CPD \cite{lebedev2014speeding} and two types of 3D SR network accelerated by separated filters. Based on three different deep learning backbones, the experimental results demonstrate that the last version of the proposed lightweight Queue module outperforms other lightweight architectures such as the lightweight modules in Xception \cite{hui2018fast} and  MobileNetV2 \cite{sandler2018mobilenetv2}.

The remainder of the paper is organized as follows. Section II introduces and summarizes work related to deep learning-based SR, typical efficient CNNs, and medical image SR. In Section III, we introduce the proposed 3D backbone model. In Section IV, several lightweight modules are presented in detail. Section V discusses our experiments and compares our results to those obtained by several state-of-the-art networks. Finally, the paper is concluded in Section VI.

\section{Related Work}

\subsection{Deep Learning-based Super-resolution}
In 2014, Dong et al. \cite{dong2015image} proposed a neural network model comprising three convolutional layers referred to as SR convolutional neural networks (SRCNN), on which the SR CNN model consisting of three parts: feature extraction, non-linear mapping, and reconstruction, is based. After Dong et al. built the basic structure of the SR network, other studies attempted to learn only the residuals between two images \cite{tai2017memnet, tai2017image, hui2018fast}. For example, Kim et al. \cite{kim2016accurate} proposed a very deep SR network that only the residual map between LR and HR images needed to be learned to restore lost high-frequency details. Most regions' residuals are close to zero; therefore, the model complexity and learning difficulty were reduced significantly.

The residual learning in ResNet \cite{he2016deep} proposed by He et al., which represents a typical local residual learning method, alleviated the degradation problem. Motivated by local residual learning, several methods based on ResNet have been proposed for image SR \cite{zhang2018image, mao2016image, han2018image, li2018multi}, which showed that the residual connection models with deep CNN structure yield high accuracy.

In addition, the dense connection is another CNN structure that is an application of the computer vision tasks \cite{zhang2018residual, tong2017image}. Based on the dense connection proposed by Huang et al. \cite{huang2017densely}, Tong et al. \cite{tong2017image} utilized several dense blocks and inserted dense connections between different dense blocks to construct a 69-layer CNN called SRDenseNet. The dense connection supports feature reuse and enhances signal propagation; thus, it has been widely used in various CNNs. Zhang et al. \cite{zhang2018residual} combined the advantages of residual connection and dense skip connection to propose a residual dense network.

In addition, various network design strategies (e.g., the learning-based up-sampling \cite{shi2016real, ledig2017photo, zhang2018learning, haris2018deep}, recursive learning \cite{kim2016deeply, lai2018fast}, multi-path learning \cite{lai2017deep, dahl2017pixel, ren2017image, wang2018fully}, and wavelet transformation \cite{bae2017beyond, guo2017deep, huang2017wavelet, liu2018multi} methods) have also been proposed for SR. However, few were designed for 3D SR or medical volume data restoration. Thus, even if we convert such models directly into 3D versions, it is difficult to achieve an effective trade-off between precision and complexity.

\subsection{Medical Image Super-resolution using Deep Learning}

Generally, 3D CNNs outperform 2D CNNs in image SR; however, 3D CNNs are difficult to train. As a result, in recent years, many researchers have continued to explore more accurate 2D models. In contrast, a small number of researchers have proposed 3D models for medical image SR by converting previous 2D models of natural image SR into 3D models. In the following, we introduce several state-of-the-art 2D and 3D CNNs for medical image SR.

\textbf{2D CNNs for Medical Image SR.} Zhu et al. \cite{zhu2019can} proposed an advanced multi-scale GAN with a lesion focused approach to achieve a stable and efficient training procedure and improved perceptual quality in super-resolved results. Zhao et al. \cite{zhao2019channel} introduced a deep channel splitting network (CSN) to ease the representational burden of deep SR models. In the CSN, feature maps are split into two different parts (i.e., branches) with different information transmissions. In addition, Zhao et al. use merge-and-run mapping \cite{zhao2016deep, hu2018single} to facilitate the integration of information from different branches. You et al. \cite{you2019ct} presented a GAN constrained by the identical, residual, and cycle learning ensemble. Du et al. \cite{du2020super} proposed a CNN with residual connections for SR reconstruction of single anisotropic 3D MR images, and Lyu et al. \cite{lyu2020mri} proposed a GAN-based deep ensemble learning framework to integrate complementary SR results from individual GAN models.

\textbf{3D CNNs for Medical Image SR.} Pham et al. \cite{pham2017brain} showed that medical volumetric data SR using 3D SRCNN outperforms slice-by-slice 2D methods; however, the restricted number of layers and simple construction limit the learning ability of 3D CNN models. Converting state-of-the-art deep 2D SR models into 3D versions and improving some structures are the most common methods to achieve better accuracy. For example, Du et al. \cite{du2018accelerated} proposed a 3D CNN using skip connections and improved dense blocks to accelerate MR image reconstruction. However, training and testing 3D models can be challenging because such models have large numbers of parameters, consume more memory, and are computationally expensive. DCSRN \cite{chen2018brain} and multi-level DCSRN \cite{chen2018efficient} proposed by Chen et al. are visually good models with relatively few parameters; however, these dense connection-based methods treat features on different channels equivalently, which does not help the models fully map the extracted features in the deep layer. In addition, GAN-based models tend to produce fake details, which is a critical problem in the medical image processing context.

\subsection{Efficient Convolutional Neural Network}

Generally, GPU-type units and conventional CNNs are available for practical applications, from embedded features in smartphones and wearable devices to robotics and autonomous driving. However, practical time constraints are often extremely stringent \cite{lebedev2018speeding}; thus, methods that use agile architecture designs have been proposed to speed up and compress CNN models \cite{chollet2017xception, jin2014flattened, szegedy2015going, szegedy2016inception,  zhang2018shufflenet, freeman2018effnet, wu2018shift}.

Determined by tensor decomposition, Jin et al. \cite{jin2014flattened} proposed a flattened CNN that uses 1D kernels in different directions to replace standard 3D convolution. In their model, Lin et al. \cite{lin2013network} replaced non-linearities in the convolutional network, with a more complex function (i.e., the multi-layer perceptron) and proposed a network-in-network architecture to reduce computational complexity. In addition, Iandola et al. \cite{iandola2016squeezenet} proposed SqueezeNet, which achieved AlexNet-level performance \cite{krizhevsky2012imagenet} with 50\% fewer parameters using smaller filters and shrinking the input channels. The MobileNet architecture proposed by Howard et al. \cite{howard2017mobilenets} separates the filtering and feature construction functions of a convolutional layer into a combination of several depthwise and pointwise convolutions. This technique was popularized due to its successful design \cite{xie2017aggregated}. Recently, separating the standard convolution layer into depthwise convolution and pointwise layers has become a research focus, and, as a result, various efficient networks frequently have been proposed, e.g., Xception \cite{chollet2017xception}, Inception \cite{szegedy2015going}, Inception-v4 \cite{szegedy2016inception}, ShuffleNets \cite{zhang2018shufflenet}, EffNet \cite{freeman2018effnet}, and ShiftNets \cite{wu2018shift}.

Historically, most research into architectures has been driven by the need to increase classification accuracy. In addition, lightweight structures based on depthwise convolution tend to cause varying degrees of accuracy degradation because the flow of information between channels is neglected. Overall, to the best of our knowledge, only a few specific and efficient 3D CNN models that can be applied to medical volumetric data SR, has been proposed to date. Thus, in this paper, to demonstrate the practical possibilities of high-efficiency networks in SR studies, we compare existing state-of-the-art methods to the proposed methods.


\section{ParallelNet}

DenseNet can integrate features of various scales in the structure; thus, it is widely used as the basic module of 3D medical volumetric data SR models for feature extraction and fusion. However, complex connections and constructions generate a significant number of parameters. Therefore, the same number of channels is frequently set for each layer of a dense block; however, feature maps have larger receptive fields in deeper layers. As a result, more channels are required for extraction and fusion. Thus, we employ the group convolution and feature aggregation to widen our model and propose a parallel connection module.

\begin{figure*}[!t]
	\centering
	\clearcaptionsetup{figure}
	\clearcaptionsetup{subfloat}
	\captionsetup[subfloat]{labelsep=none,format=plain,labelformat=empty}
	\subfloat[(a)]{\includegraphics[width=4.5in]{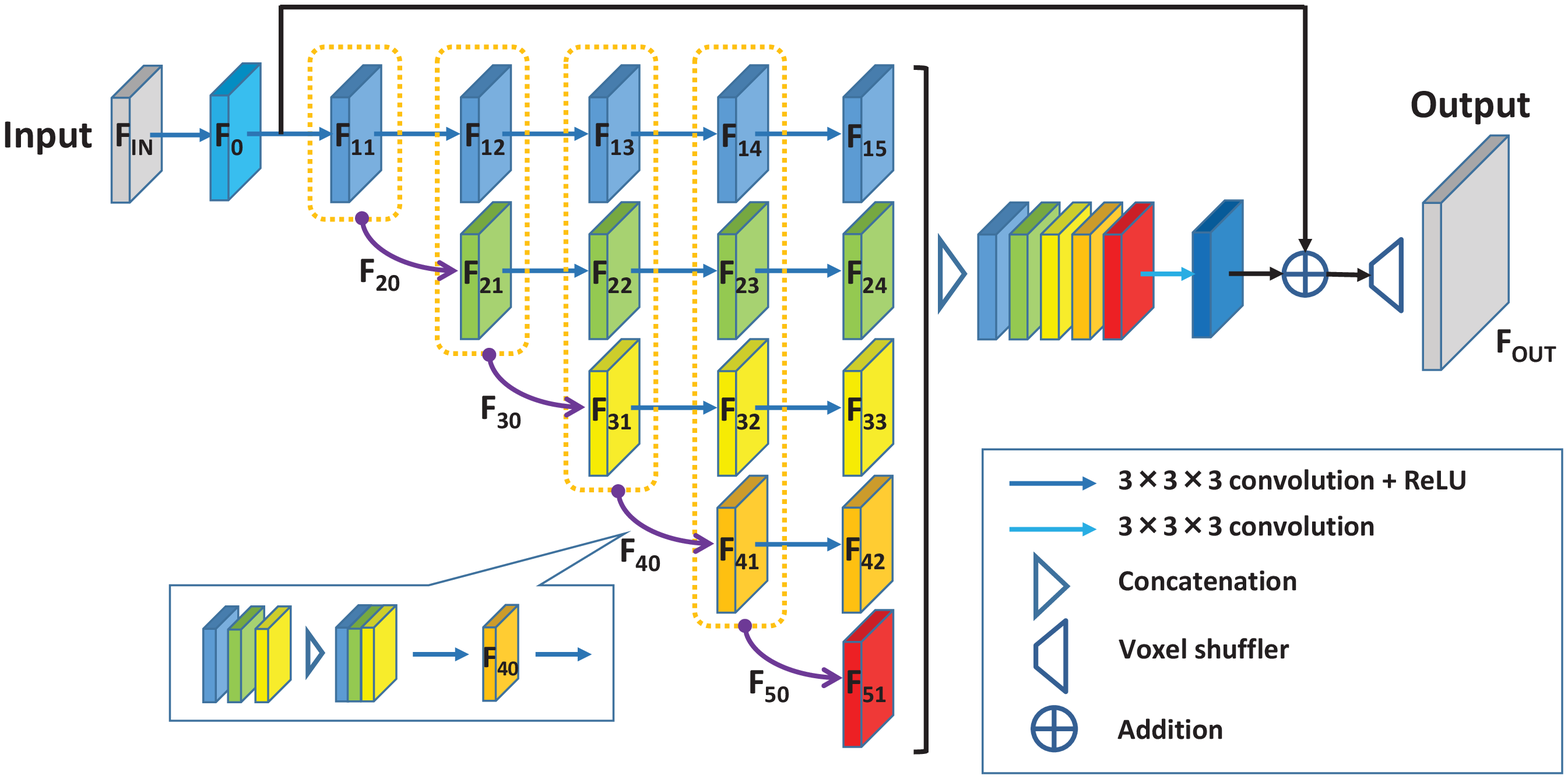}}
  \hfil
  \subfloat[(b)]{\includegraphics[width=2in]{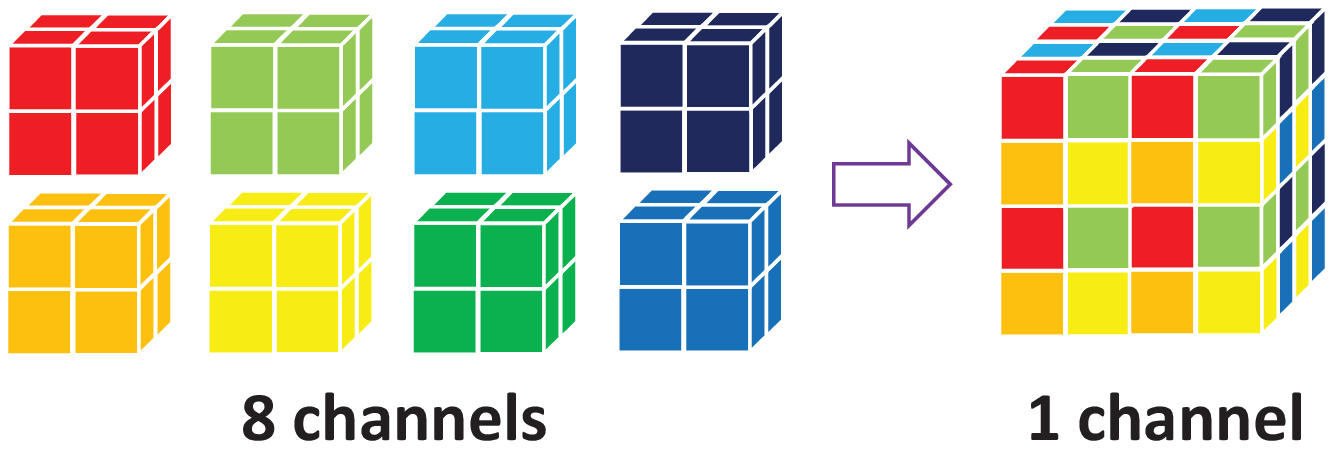}}
	\vfil
	\subfloat[(c)]{\includegraphics[width=4.5in]{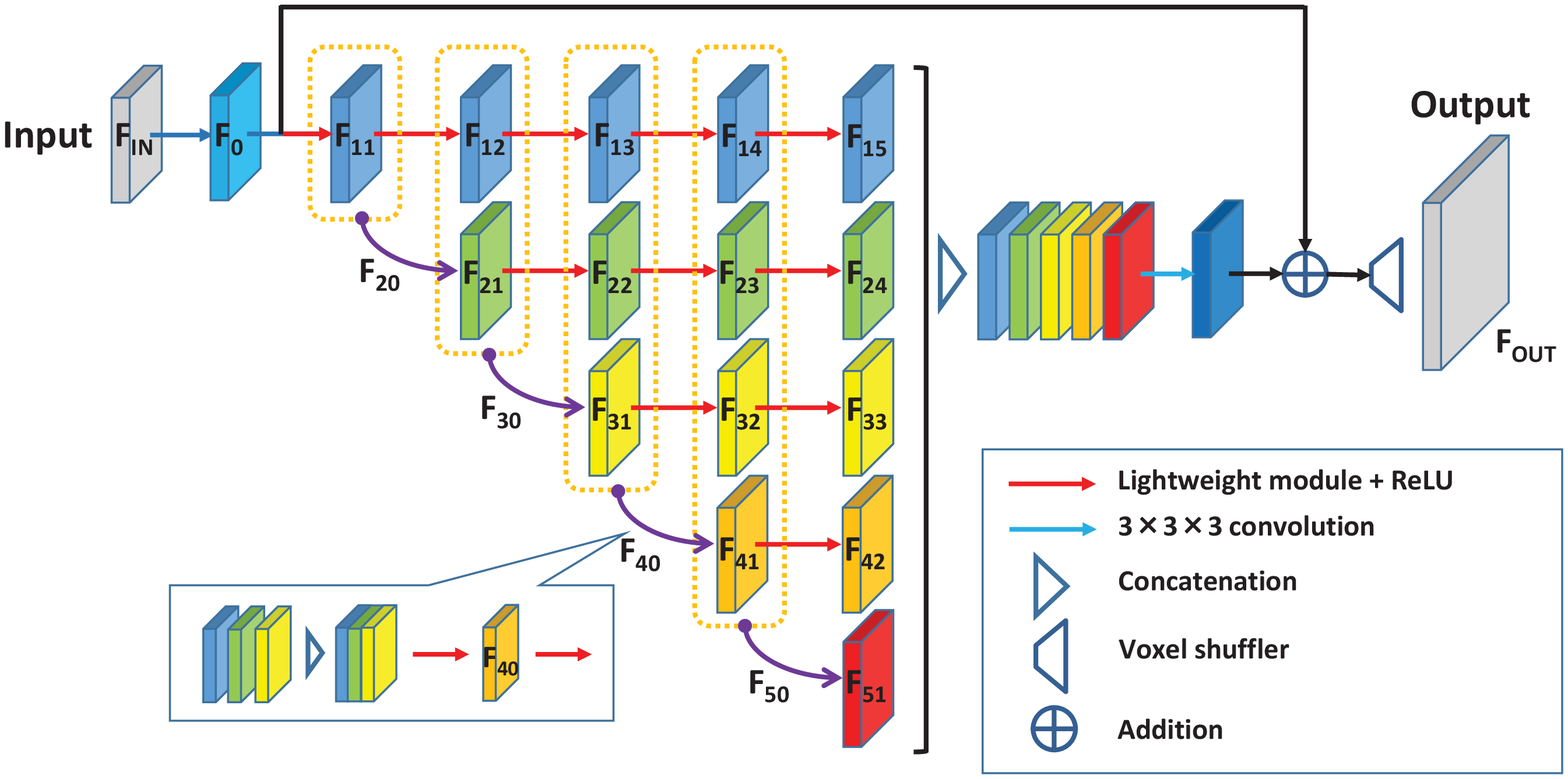}}
	\caption{(a) ParallelNet, (b) Voxel shuffle operation, and (c) VolumeNet. Here, we use ParallelNet with 5 layers in the mainline (e.t., the first branch in (a)) as an example. In our proposed lightweight model, the convolutional layers can be replaced by lightweight modules (illustrated in Fig. 2(b)-(g)), except for the first and last convolutional layers.}
	\label{fig_1}
\end{figure*}

As shown in Fig. 1(a), each branch (i.e., the convolutional layers arranged horizontally in the parallel connection module) can be considered to be separated from the mainline (i.e., the first branch) using group convolution. In addition, to better integrate the feature maps of other branches, we employ feature aggregation modules, where all feature maps of the previous layer of other branches are aggregated as the input to a new branch. The proposed ParallelNet primarily comprises feature extraction, a parallel connection module, global residual connection, and up-sampling. In the following, we introduce each component in detail.

\subsection{Feature Extraction}

Here, the input and output of ParallelNet are denoted \(\mathbf{F}_{\mathrm{IN}}\) and \(\mathbf{F}_{\mathrm{OUT}}\), respectively. Note that we employ a single convolution layer to extract shallow features. The first convolution layer extracting features \(\mathbf{F}_{0}\) from the LR input are obtained as follows:
\begin{equation}
\begin{aligned}
	\mathbf{F}_{0}=H_{\mathrm{EX}}(\mathbf{F}_{\mathrm{IN}}),
\end{aligned}
\end{equation}
where \(H_{\mathrm{EX}}(\cdot)\) denotes the convolution operation. Then, \(\mathbf{F}_{0}\) is used for global residual connection.

\subsection{Parallel Connection Module}

Assume we have \(D\) layers in the mainline (first branch) of the parallel connection module. Then, the output \(\mathbf{F}_{i,j}\) \((j \geq 1)\) of the \(j\)-th layer in the \(i\)-th line can be obtained as follows:
\begin{equation}
\begin{aligned}
	\mathbf{F}_{i,j}=\sigma (H_{i,j}(\mathbf{F}_{i,j-1})),
\end{aligned}
\end{equation}
where \(H_{i,j}(\cdot)\) denotes the convolution operation of the \(j\)-th layer in the \(i\)-th branch and \(\sigma\) denotes rectified linear units (ReLU) \cite{glorot2011deep}.

To fully extract and fuse features, as the core component of the proposed ParallelNet, we present a feature aggregation module that combines feature maps with the same receptive fields in other layers at the beginning of each branch line. Thus, the input of the first layer of each branch \(\mathbf{F}_{i,0}\) can be expressed as follows:
\begin{equation}
\begin{aligned}
	\mathbf{F}_{i,0}=\left\{\begin{matrix}
	\mathbf{F}_{0} & i=1 \\ 
	\sigma (H_{i,0}(\left [ \mathbf{F}_{1,i-1},\mathbf{F}_{2,i-2},\dots,\mathbf{F}_{i-1,1} \right ])) & 2\leq i\leq D
	\end{matrix}\right.,
\end{aligned}
\end{equation}
where \([\mathbf{F}_{1,i-1},\mathbf{F}_{2,i-2},\dots,\mathbf{F}_{i-1,1}]\) represents the concatenation of the feature maps produced by the \((i-1)\)-th convolutional layer in the first branch (mainline), the \((i-2)\)-th convolutional layer in the second branch,…, and the first convolutional layer in the \((i-1)\)-th branch, which results in \(DK\) feature maps. In our SR model, the number of channels in each layer is \(K\). Here, \(H_{i,0}(\cdot)\) denotes the convolution operation, which shrinks the channel from \((i-1)K\) to \(K\) and preliminarily fuses the feature maps obtained from the other branches. \(\sigma\) denotes the ReLU activation function.

To fully exploit the hierarchical features extracted by the parallel connection module, we conduct feature aggregation using a concatenation layer and a convolution layer. Here, the convolutional layer fully fuses the extracted features prior to fusion, and the convolutional layer squeezes channels after concatenation to make the two input terminals of the global residual layer the same size. We refer to this operation as feature fusion, which is expressed as follows:
\begin{equation}
\begin{aligned}
  \mathbf{F}_{\mathrm{PC}}=H_{\mathrm{PC}}([\mathbf{F}_{1,D},\mathbf{F}_{2,D-1},\dots,\mathbf{F}_{D,1}]),
\end{aligned}
\end{equation}
where \([\mathbf{F}_{1,D},\mathbf{F}_{2,D-1},\dots,\mathbf{F}_{D,1}]\) represents the concatenation of the feature maps produced by the outputs of the mainline and branch two, \dots, \(D\), and \(H_{\mathrm{PC}}(\cdot)\) denotes the convolution function.

\subsection{Global Residual Connection}

The global residual connection can effectively avoid gradient vanishing problem, reduce training difficulty, and increase training speed; therefore, we employ a global residual connection structure.

The output of the global residual layer \(\mathbf{F}_{\mathrm{GRN}}\) can be expressed as follows:
\begin{equation}
\begin{aligned}
  \mathbf{F}_{\mathrm{GRN}}=\mathbf{F}_{0} \oplus \mathbf{F}_{\mathrm{PC}}
\end{aligned}
\end{equation}
where \(\oplus\) denotes the matrix addition operation.

\subsection{Up-sampling}

Inspired by previous studies \cite{shi2016real, zhang2018residual}, we employ a voxel shuffle layer as the up-sampling net, which rapidly reconstructs the HR image without parameters (Fig. 1(b)). In addition, using the voxel shuffle layer as the final layer of the entire model helps the model learn the subvoxel features of HR images. The voxel shuffle operation is denoted as \(VS(\cdot)\), and the output of ParallelNet is expressed as follows:
\begin{equation}
\begin{aligned}
  \mathbf{F}_{\mathrm{OUT}}=VS(\mathbf{F}_{\mathrm{GRN}}).
\end{aligned}
\end{equation}

\section{Lightweight ParallelNet}

To further improve the accuracy and efficiency of the proposed ParallelNet based on standard full 3D convolution (Fig. 2(a)), here, we present the lightweight Queue module. The proposed Queue was inspired by Xception \cite{hui2018fast} and MobileNetV2 \cite{sandler2018mobilenetv2}, as shown in Fig. 2(b) and Fig. 2(c), respectively.

\begin{figure*}[!t]
  \centering
  \includegraphics[width=7in]{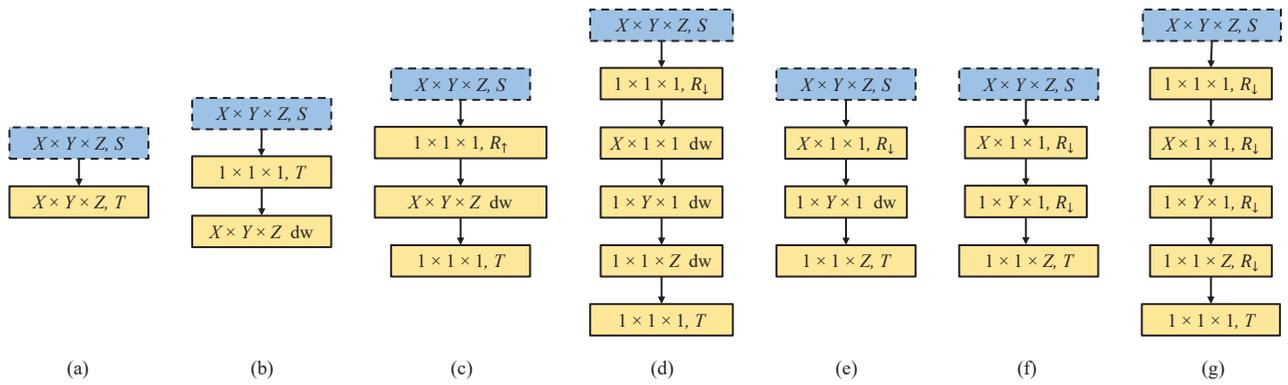}
  \caption{(a) Standard 3D full convolution, (b) Xception module (3D ver.), (c) MobileNetV2 module (3D ver.), (d) CPD module (3D ver.), (e) LWv1 module, (f) LWv2 module, and (g) Queue module. Here, dw represents the depthwise operation. \(R_{\uparrow}\) and \(R_{\downarrow}\) are the number of convolution kernels for expanding and shrinking input channels of the module (marked in blue), respectively.}
  \label{fig_2}
\end{figure*}

The combination of depthwise and pointwise convolutions greatly reduces the number of network parameters, accelerates network prediction, and avoids overfitting. However, due to insufficient integration between channels, the depthwise convolution-based lightweight structure tends to reduce network accuracy. Thus, differing from existing lightweight models that use depthwise convolutions (Fig. 2(b) and Fig. 2(c)), we apply cross-channel convolutions on the combination of depthwise and pointwise convolutions layers (Fig. 2(g)). This structure plays the role of standard convolutions and ensures that the information of all feature maps is merged in each step, which ensures that accuracy is not reduced. We refer to this structure as the Queue module, and we embed Queue models into the proposed ParallelNet for higher accuracy and efficiency (i.e., VolumeNet). To facilitate comparison, we demonstrate a 3D SR network accelerated by CPD \cite{lebedev2014speeding} (Fig. 2(d)) and two 3D SR networks accelerated by separated filters, as shown in Figs. 2(e) and 2(f).

\subsection{Standard 3D Convolution}

Generally, weights in 2D CNNs can be described by 3D filters as \(\mathbf{W}\in \mathbb{R}^{X\times Y\times S}\) where \textit{S} is the number of input channels, \textit{X} and \textit{Y} are the spatial dimensions of the filter. Specifically, the convolution for each output channel can be expressed as follows:
\begin{equation}
\begin{aligned}
  \mathbf{F}_{t}(x,y)=&\sum_{{x}'=-X/2}^{X/2}\sum_{{y}'=-Y/2}^{Y/2}\sum_{s=1}^{S}\\
  &\mathbf{F}_{\mathrm{2D}}(x-{x}',y-{y}',s)\mathbf{W}_{t}({x}',{y}',s),
\end{aligned}
\end{equation}
where \(\mathbf{F}_{\mathrm{2D}}\in \mathbb{R}^{H\times W\times S} \) is a set of 2D input feature maps with a size of \(H\times W\) and t is the output channel index.

Weights in standard 3D CNNs can be described by four-dimensional filters as \(\mathbf{W}\in \mathbb{R}^{X\times Y\times Z\times S}\) where \textit{S} is the number of input channels and \textit{X}, \textit{Y}, and \textit{Z} are the spatial dimensions of the filter. In previous SR models, the standard 3D convolution shown in Fig. 2(a) can be generally expressed as follows:
\begin{equation}
\begin{aligned}
  \mathbf{F}_{t}(x,y,z)=&\sum_{{x}'=-X/2}^{X/2}\sum_{{y}'=-Y/2}^{Y/2}\sum_{{z}'=-Z/2}^{Z/2}\sum_{s=1}^{S}\\
&\mathbf{F}_{\mathrm{3D}}(x-{x}',y-{y}',z-{z}',s)\mathbf{W}_{t}({x}',{y}',{z}',s),
\end{aligned}
\end{equation}
where \(\mathbf{F}_{\mathrm{3D}}\in \mathbb{R}^{H\times W\times L\times S}\) is a set of 3D output images of the previous layer, \textit{t} is the output channel index, and \textit{H}, \textit{W}, and \textit{L} are the spatial dimensions of the input.

The training parameters of a 2D convolution layer are \(X \times Y \times S \times T\), where \(X \times Y\) represents the kernel size, \(S\) is the number of input channels, and \(T\) is the output channels. Thus, the parameters of the corresponding 3D convolution layer will be \(X \times Y \times Z \times S \times T\), i.e., the number of parameters will increase \(Z\) times when a 2D network is converted to a 3D model. Therefore, constructing highly efficient and accurate 3D CNN models SR for medical volumetric data is challenging.

\subsection{CPD based Lightweight Module}

Here, to minimize the number of parameters, we present a lightweight model based on CPD \cite{lebedev2014speeding} for medical volumetric image SR. The CPD factorizes a tensor into a sum of outer products of vectors, e.g., for a three-way tensor \(\mathbf{P}\), the CP decomposition can be expressed as follows:
\begin{equation}
\begin{aligned}
  \mathbf{P}\approx \sum_{r=1}^{\mathfrak{R}}\mathbf{u}_{r}\circ \mathbf{v}_{r}\circ \mathbf{w}_{r},
\end{aligned}
\end{equation}
where \(\mathfrak{R}> 0\), and \(\mathbf{u}_{r}\), \(\mathbf{v}_{r}\), \(\mathbf{w}_{r}\) are vectors of appropriate dimensions, and ``\(\circ\)'' denotes the outer product for tensors, i.e.,
\begin{equation}
\begin{aligned}
  {p}_{ijk} \approx \sum_{r=1}^{\mathfrak{R}}\mathbf{u}_{ri}\circ \mathbf{v}_{rj}\circ \mathbf{w}_{rk}.
\end{aligned}
\end{equation}

Under the rank-one assumption of CPD (i.e., \(\mathfrak{R}=1\)), the filter \(\mathbf{\hat{W}}\in \mathbb{R}^{X\times Y\times Z\times S}\) will be separated into cross-products of four 1D filters as follows:
\begin{equation}
\begin{aligned}
  \mathbf{\hat{W}}_{t}=\bm{\alpha}_{t}\times\bm{\beta}_{t}\times\bm{\gamma}_{t}\times\bm{\delta}_{t},
\end{aligned}
\end{equation}
where \(\bm{\alpha}_{t}\), \(\bm{\beta}_{t}\), \(\bm{\gamma}_{t}\), and \(\bm{\delta}_{t}\) are 1D convolution vectors convolving features across the X, Y, and Z dimensions, and channels, respectively. Substituting Equation 11 into Equation 8, the 3D convolution layer is modified as follows:
\begin{equation}
\begin{aligned}
  &\mathbf{\hat{F}}_{t}(x,y,z)=\mathbf{F}_{\mathrm{3D}}\ast\mathbf{\hat{W}}_{t}=\sum_{s=1}^{R}(\sum_{{z}'=-Z/2}^{Z/2}(\sum_{{y}'=-Y/2}^{Y/2}(\sum_{{x}'=-X/2}^{X/2}\\
  &\mathbf{F}_{\mathrm{3D}}(x-{x}',y-{y}',z-{z}',s)\bm{\alpha}_{t}({x}'))\bm{\beta}_{t}({y}'))\bm{\gamma}_{t}({z}'))\bm{\delta}_{t}(s),
\end{aligned}
\end{equation}
where \(\mathbf{F}_{\mathrm{3D}}\in \mathbb{R}^{H\times W\times L\times R}\) is the output of the first layer in the CPD module and \(\ast\) represents the convolution operation. Fig. 2(d) illustrates the 3D convolution operations of a fully decomposed layer through CPD.

Here, we present a simple comparison. A standard full convolution (Fig. 2(a)) was inserted into the 3D CNNs, where the numbers of channels (3D feature maps) before and after an operation are \(S\) and \(T\), respectively. In addition, the kernel size is \(X \times Y \times Z\), and the operation in the convolution layer requires \(X \times Y \times Z \times S \times T\) parameters. As shown in Fig. 2(b), the decomposed 3D convolution initially reduces the feature maps from \(S\) to \(R\) by \(1 \times 1 \times 1\) convolution. Then, three 1D convolutions from three axes (i.e., \(X \times 1 \times 1\), \(1 \times Y \times 1\), and \(1 \times 1 \times Z\)) are applied to convolve features in each channel followed by the \(1 \times 1 \times 1\) convolution to increase the channel from \(R\) to \(T\). Thus, the operation shown in Fig. 2(b) requires \(SR + (X + Y + Z) R + RT\) parameters. With CPD, the compression ratio \(E_{\mathrm{CPD}}\) is given as follows:
\begin{equation}
  E_{\mathrm{CPD}}=\frac{XYZST}{SR+(X+Y+Z)R+RT}.
\end{equation}

\subsection{Lightweight Module-v1 (LWv1) and Lightweight Module-v2 (LWv2)}

Since CPD-based 3D convolution tends to reduce the accuracy of the network, we attempt to convert a two-component architecture \cite{rigamonti2013learning} to a 3D version. Thus, here, we demonstrate two lightweight modules, i.e., Lightweight Module-v1 (LWv1) and Lightweight Module-v2 (LWv2), as shown in Figs. 2(e) and 2(f), respectively.

The convolution for the first output layer of LWv1 and LWv2 can be expressed as follows:
\begin{equation}
\begin{aligned}
  \mathbf{F}_{t}^{(1)}(x,y,z)=\sum_{s=1}^{R}\sum_{{x}'=-X/2}^{X/2}\mathbf{F}_{\mathrm{3D}}(x-{x}',y,z,s)\bm{\tilde{\alpha}}_{t}({x}',s),
\end{aligned}
\end{equation}
where \(\bm{\tilde{\alpha}}_{t}({x}',s)\) is the 2D convolution kernel across the channels and X dimension.

Note that the main difference between LWv1 and LWv2 is the convolution of the middle layer. The convolution of LWv1 for the second output layer can be expressed as follows:
\begin{equation}
\begin{aligned}
  \mathbf{F}_{t}^{(2)}(x,y,z)=\sum_{{y}'=-Y/2}^{Y/2}\mathbf{F}_{t}^{(1)}(x,y-{y}',z)\bm{\tilde{\beta}}_{t}({y}',1),
\end{aligned}
\end{equation}
where \(\bm{\tilde{\beta}}_{t}({y}',1)\) represents the 1D convolution kernels operated in the \(t\)-th channel and Y dimension.

When we add a pointwise convolution operation after Equation 15, we obtain the following:
\begin{equation}
\begin{aligned}
  {\mathbf{F}_{t}}'^{(2)}(x,y,z)=\sum_{s=1}^{R}\sum_{{y}'=-Y/2}^{Y/2}\mathbf{F}_{t}^{(1)}(x,y-{y}',z,s)\bm{\tilde{\beta}}_{t}({y}',s),
\end{aligned}
\end{equation}
where \(\bm{\tilde{\beta}}_{t}({y}',s)\) is the 2D convolution kernel across the channels and Y dimension. Note that this represents the second output channel of LWv2.

In the third layer, we perform channel expansion and convolution in the Z direction simultaneously, and this operation is expressed as follows:
\begin{equation}
\begin{aligned}
  \mathbf{F}_{t}^{(3)}(x,y,z)=\sum_{s=1}^{T}\sum_{{z}'=-Z/2}^{Z/2}\mathbf{F}_{t}^{(2)}(x,y,z-{z}',s)\bm{\tilde{\gamma}}_{t}({z}',s).
\end{aligned}
\end{equation}
where \(\bm{\tilde{\gamma}}_{t}({z}',s)\) is the 2D convolution kernel across the channels and Z dimension.

In the case shown in Fig. 2(e), 2D convolution with kernel size \(X \times 1 \times 1\) is performed in the first layer followed by depthwise convolution using \(1 \times Y \times 1\) filters. Then, 2D convolution with kernel size \(1 \times 1 \times Z\) is performed in the final layer. Thus, LWv1 requires \(XSR + YR + ZRT\) parameters. The compression ratio \(E_{\mathrm{LWv1}}\) is expressed as follows:
\begin{equation}
\begin{aligned}
  E_{\mathrm{LWv1}}=\frac{XYZST}{XSR+YR+ZRT}.
\end{aligned}
\end{equation}

Differing from LWv1, in the middle layer of LWv2, we utilize the 2D convolution with kernel size \(1 \times Y \times 1\) rather than depthwise convolution. Thus, operation by this model requires \(XSR + YR^{2} + ZRT\) parameters and the compression ratio \(E_{\mathrm{LWv2}}\) is expressed as follows:
\begin{equation}
\begin{aligned}
  E_{\mathrm{LWv2}}=\frac{XYZST}{XSR + YR^{2} + ZRT}.
\end{aligned}
\end{equation}


\subsection{Queue Module}

We found that LWv2 only adds a single hidden pointwise convolution structure into LWv1; however, the accuracy is improved. Therefore, we attempt to change all depthwise layers to full channel convolution. In addition, inspired by Xception \cite{hui2018fast} and MobileNetV2 \cite{sandler2018mobilenetv2} (Figs. 2(b) and 2(c), respectively), we employ channel shrinking and expansion to implement mapping in a low-dimensional space and propose the final lightweight model, i.e., the proposed Queue module, as shown in Fig. 2(g). Here, we combine the textures in all channels in each step; thus, there is no accuracy loss when shrinking channels. In the proposed Queue module, each output channel in the second, third, and fourth layers can be expressed as follows:
\begin{equation}
  \mathbf{F}_{t}^{\bm{\alpha}}(x,y,z)=\sum_{s=1}^{R}\sum_{{x}'=-X/2}^{X/2}\mathbf{F}_{\mathrm{3D}}(x-{x}',y,z,s)\bm{\tilde{\alpha}}_{t}({x}',s),
\end{equation}
\begin{equation}
  \mathbf{F}_{t}^{\bm{\beta}}(x,y,z)=\sum_{s=1}^{R}\sum_{{y}'=-Y/2}^{Y/2}\mathbf{F}^{\bm{\alpha}}(x,y-{y}',z,s)\bm{\tilde{\beta}}_{t}({y}',s),
\end{equation}
\begin{equation}
  \mathbf{F}_{t}^{\bm{\gamma}}(x,y,z)=\sum_{s=1}^{R}\sum_{{z}'=-Z/2}^{Z/2}\mathbf{F}^{\bm{\beta}}(x,y,z-{z}',s)\bm{\tilde{\gamma}}_{t}({z}',s),
\end{equation}
where \(\bm{\tilde{\alpha}}_{t}\), \(\bm{\tilde{\beta}}_{t}\), and \(\bm{\tilde{\gamma}}_{t}\) are 2D convolution kernels across the channels and X, Y, and Z dimensions, respectively. Here, \(t\) ranges from 1 to \(R\). These operations make the model always integrate the information of all feature maps from each local direction.

As shown in Fig. 2(g) shows, the decomposed 3D convolution initially reduces the channel from \(S\) to \(R\) via the \(1 \times 1 \times 1\) convolution, and then, three 1D convolutions from three axes (i.e., \(X \times 1 \times 1\), \(1 \times Y \times 1\), and 1 × 1 × Z) are applied to convolve features across all channels. In the final layer, we employ \(1 \times 1 \times 1\) convolution to increase the channel from \(R\) to \(T\). As a result, the operation shown in Fig. 2(g) requires \(SR + (X + Y + Z) R^{2} + RT\) parameters. With the proposed Queue module, the compression ratio \(E_{\mathrm{Queue}}\) is expressed as follows:
\begin{equation}
  E_{\mathrm{Queue}}=\frac{XYZST}{SR + (X + Y + Z) R^{2} + RT}.
\end{equation}

Our experimental results demonstrate that this structure can improve accuracy compared to the undecomposed model (Section V).

\subsection{VolumeNet}

To ensure the integrity of the extracted or reconstructed feature maps as much as possible, we retain the standard convolution in the first and last layers of the original ParallelNet. Note that a convolutional layer whose input or output feature map is 1 cannot be compressed; thus, we only replace the specific standard 3D convolution layers by the proposed Queue modules in ParallelNet and present a more accurate and efficient ParallelNet named VolumeNet, as shown in Fig. 1(c).

\subsection{Loss Function}

In this study, our models are optimized by the L1 norm function. The absolute difference (L1 norm) between a network’s output SR and ground truth (GND) HR images can be expressed by Equation 24.
\begin{equation}
  Loss_{\mathrm{L1}}=\frac{1}{HWL}\sum_{z=1}^{L}\sum_{y=1}^{W}\sum_{x=1}^{H}\left | \mathbf{I}_{x,y,z}^{\mathrm{SR}}-\mathbf{I}_{x,y,z}^{\mathrm{HR}} \right |,
\end{equation}
where \(\mathbf{I}_{x,y,z}^{\mathrm{SR}}\) is the SR output from the deep learning model, and \(\mathbf{I}_{x,y,z}^{\mathrm{HR}}\) is the GND HR image of size \(H \times W \times L\).

Although the L2 norm is widely used in image restoration tasks due to its relationship with the peak signal-to-noise ratio (PSNR), the L1 norm depicts better convergence and performance \cite{zhang2018residual, ahn2018fast, lim2017enhanced, zhao2016loss}. Note that the drawback of the L1 norm is that convergence speed is relatively slower compared to L2 norm without the residual block. However, utilizing a residual learning model could mitigate this drawback effectively.

\begin{figure}[t]
  \centering
  \includegraphics[width=3.4in]{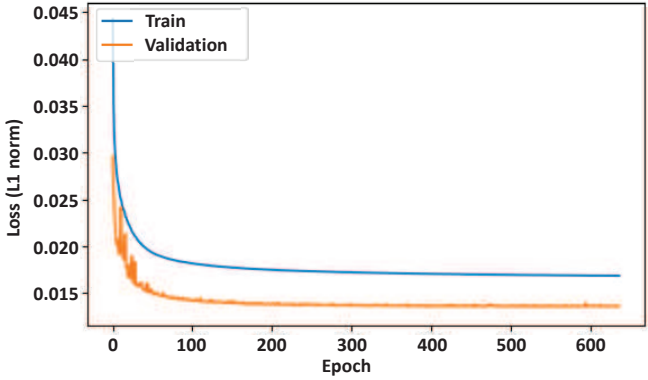}
  \caption{Convergence curve of training set and validation set on IXI database.}
  \label{fig_3}
\end{figure}

\section{Experiment}

\subsection{Data Preparation and Evaluation Protocols}

Here we provide details about our experimental settings, e.g., the training and testing data used for each task.

\textbf{Data of MRI Image SR.} We use T1 images from the IXI database \cite{ixi}, which is an extensive public database of brain MRI images. The IXI database includes T1-weighted images of 544 healthy subjects, where the subject age range is 20 to 86 years old. The data were collected at three different hospitals in London using 1.5T and 3T scanners. To prove that the proposed method is suitable for processing various types of data, we mixed the original data and randomly selected 50 training, 10 validation, and 100 test data of size \(250 \times 250 \times 150\) voxels to compare the quality of our networks. These data were used as GND HR images and degraded to LR images at a \(\times2\) scale in the X, Y, and Z directions via tricubic interpolation.

\textbf{Training Setting.} We implemented all models in Keras with an NVIDIA Quadro RTX 8000 GPU. The operating system and central processing unit were Ubuntu 16.04 LTS and an Intel Core i9-9820X, respectively.

We randomly extracted eight LR patches of size \(16 \times 16 \times 16\) voxels as inputs in each training batch of the proposed ParallelNet, VolumeNet, and other derived lightweight versions. The other state-of-the-art models followed the settings found in the corresponding literature.

Here, we used the Adam optimizer \cite{kingma2014adam} and set the learning rate to \(10^{-4}\). We set the number of epochs to infinity, and training was terminated when the loss function had 50 subsequent epochs without reduction. Although there was no obvious overfitting according to each model's recorded convergence value during training, we only kept the weight when the model achieved the best accuracy to avoid possible overfitting. Fig. 3 shows the training curve with validation of the proposed ParallelNet on the IXI database. The total training time for ParallelNet was approximately seven hours.

\begin{figure}[t]
  \centering
  \includegraphics[width=3.4in]{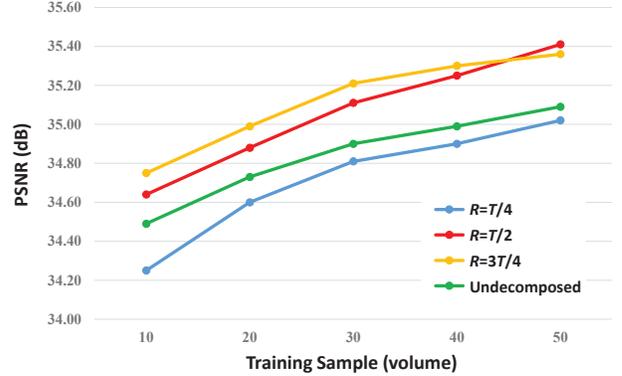}
  \caption{Accuracy and training samples of (undecomposed) ParallelNet and VolumeNets on IXI database. The undecomposed one represents the standard ParallelNet and the others are VolumeNet with different compression ratios of channels in the Queue module.}
  \label{fig_4}
\end{figure}

\subsection{Network Investigations on IXI database}

\textbf{Study of depth and width.} We investigated the basic network parameters, i.e., the number of layers in the mainline of a parallel module (denoted \(D\)) and the number of channels in convolution layers (denoted \(C\)) except the first layer. Note that the number of channels in the first layer must be 8 due to our adoption of the voxel shuffle layer. In traditional cognition, a network with greater \(D\) and greater \(C\) becomes deeper and wider, thereby obtaining better performance. However, in the case of limited training samples, when \(D\) is greater than 5, the network is easy to overfit. Thus, its learning ability will decline (Table I). In this case, using lightweight modules rather than standard convolutions can reduce the number of parameters and prevent overfitting effectively. In addition, by using lightweight modules with a smart structure, we can design deeper and wider models to achieve better accuracy.

\begin{table*}[!t]
\renewcommand{\arraystretch}{1.5}
\caption{PSNR of models with \(C\) channels per layer (except for the first layer) and \(D\) layers in the mainline of a parallel connection module on MRI dataset.}
\label{table_1}
\centering
\begin{tabular}{cccccccccc}
  \hline \hline
  \(C \backslash D\) & 2 & 3 & 4 & 5 & 6 & 7 & 8 & 9 & 10\\
  \hline
  1 & 32.33 & \cellcolor[HTML]{FAFCFE}32.55 & \cellcolor[HTML]{F6FAFD}32.70 & \cellcolor[HTML]{F4F8FD}32.80 & \cellcolor[HTML]{F2F7FC}32.89 & \cellcolor[HTML]{F0F6FB}32.97 & \cellcolor[HTML]{EDF5FB}33.06 & \cellcolor[HTML]{ECF3FA}33.13 & \cellcolor[HTML]{EAF2FA}33.20 \\
  \hline
  16 & \cellcolor[HTML]{D4E5F4}34.10 & \cellcolor[HTML]{CADFF2}34.48 & \cellcolor[HTML]{C3DBF0}34.76 & \cellcolor[HTML]{BFD8EF}34.94 & \cellcolor[HTML]{ABCBE7}35.05 & \cellcolor[HTML]{88B3D9}35.15 & \cellcolor[HTML]{96BCDF}35.11 & \cellcolor[HTML]{93BADD}35.12 & \cellcolor[HTML]{8FB8DC}35.13 \\
  \hline
  32 & \cellcolor[HTML]{CBE0F2}34.44 & \cellcolor[HTML]{C0D9EF}34.88 & \cellcolor[HTML]{A4C6E4}35.07 & \cellcolor[HTML]{88B3D9}35.15 & \cellcolor[HTML]{93BADD}35.12 & \cellcolor[HTML]{9DC1E1}35.09 & \cellcolor[HTML]{96BCDF}35.11 & \cellcolor[HTML]{9DC1E1}35.09 & \cellcolor[HTML]{A0C3E3}35.08 \\
  \hline
  32 + Queue & \cellcolor[HTML]{CFE2F3}34.28 & \cellcolor[HTML]{C7DDF1}34.61 & \cellcolor[HTML]{C0D9EF}34.89 & \cellcolor[HTML]{9ABFE0}35.10 & \cellcolor[HTML]{6DA0CE}35.23 & \cellcolor[HTML]{4E8BC2}35.32 & \cellcolor[HTML]{4484BE}35.35 & \cellcolor[HTML]{2F75B5}35.41 & \cellcolor[HTML]{3378B7}35.40\\
  \hline \hline
\end{tabular}
\end{table*}

\textbf{Efficiency Trade-off.} Fig. 4 shows the proposed VolumeNet, whose compression ratio of the channel is 25\%, 50\%, or 75\%, outperformed the undecomposed ParallelNet. The compression ratio 50\% achieved a good trade-off between accuracy and complexity; therefore, we used it in all experiments.

For example, in each lightweight module of VolumeNet, the \(1 \times 1 \times 1\) convolution was adopted in the first layer to shrink channels from 32 (\(S\)) to 16 (\(R\)). Then, three convolution layers with kernel \(3 \times 1 \times 1\), \(1 \times 3 \times 1\), and \(1 \times 1 \times 3\) were set for the mapping operations. We then adopted the second \(1 \times 1 \times 1\) convolution to expand the channel from 16 (\(R\)) to 32 (\(T\)) followed by a ReLU function.

\begin{table}[!t]
\renewcommand{\arraystretch}{1.5}
\caption{Ablation investigation of global residual connection (GRC), parallel connections (PC), and Queue module (QM). We observe the performance (PSNR) on MRI images (IXI dataset T1 data) with scaling factor \(\times 2\).}
\label{table_2}
\centering
\begin{tabular}{ccccccc}
  \hline \hline
  Group	& GRC &	PC & QM &	PSNR & Time (s) &	\#Param. (M)\\
  \hline
  A &   &	  &	  &	35.15 &	0.156 &	0.236\\
  B	& 	&  	& \ding{52}	& 34.89	& 0.138	& 0.038\\
  C	& 	& \ding{52}	& 	& 35.02	& 0.902	& 2.061\\
  D	& 	& \ding{52}	& \ding{52}	& 35.23	& 0.484	& 0.230\\
  E	& \ding{52} &   &	  &	35.12 &	0.159 &	0.236\\
  F	& \ding{52}	& 	& \ding{52}	& 34.89	& 0.137	& 0.038\\
  G	& \ding{52}	& \ding{52}	& 	& 35.14	& 0.904	& 2.061\\
  H	& \ding{52}	& \ding{52}	& \ding{52}	& 35.41	& 0.492	& 0.230\\
  \hline \hline
\end{tabular}
\end{table}

\textbf{Ablation Investigation.} Table II shows the ablation investigation on the effects of global residual connection, parallel connections, and the Queue module. Here, the five networks had the same depth in the first branch (\(D=9\)) and width (\(C=32\)). The experimental results demonstrated the following three main points:

1) Global residual connection does not affect the results of shallow CNNs and improves the accuracy of deep CNNs because residual learning can effectively reduce the training difficulty of deep complex networks (refer to Groups A \& E, Groups B \& F, Groups C \& G, and Groups D \& H).

2) The parallel connections enhance the learning ability of the network and significantly increase accuracy (refer to Groups B \& D, Groups E \& G, Groups F \& H). However, using only the parallel connections without global residual connection or the Queue module will increase training difficulty due to the presence of too many parameters (refer to Groups A \& C).

3) Using the proposed Queue module directly on a simple CNN reduces the expression ability of the model due to too few parameters; thus, the effect is not good (refer to Groups A \& B, Groups E \& F). When the hierarchical feature fusion in the backbone structure is well-designed, the Queue module will significantly improve the performance and efficiency of the network (refer to Groups C \& D, Groups G \& H).

\subsection{Comparison with Other Methods on IXI database}

\begin{table}[!t]
\renewcommand{\arraystretch}{1.5}
\caption{Quantitative results of MRI images on IXI database.}
\label{table_3}
\centering
\begin{tabular}{cccccc}
\hline \hline
Method & RMSE &	PSNR & SSIM & \begin{tabular}[c]{@{}c@{}}Time\\ (s)\end{tabular} & \begin{tabular}[c]{@{}c@{}}\#param.\\ (M)\end{tabular}\\
\hline
Tricubic interpolation & 6.752 & 31.54 & 0.9745 &	- &	-\\
\hline
3D SRCNN \cite{pham2017brain} & 5.187 & 33.83 & 0.9812 & 0.338 & 0.053\\
\hline
DCSRN \cite{chen2018brain} & 4.789 & 34.53 & 0.9830 & 1.797 & 0.224\\
\hline
mDCSRN \cite{chen2018efficient} & 4.667 & 34.75 & 0.9831 & 4.341 & 0.425\\
\hline
\begin{tabular}[c]{@{}c@{}}ParallelNet (D5, C32) \\ {[}ours{]}\end{tabular} & 4.366 & 35.15 & 0.9839 & 0.307 & 0.595 \\
\begin{tabular}[c]{@{}c@{}}VolumeNet (D5, C32)\\ {[}ours{]}\end{tabular}   & 4.484 & 35.10 & 0.9839 & \textbf{0.201} & 0.089 \\
\begin{tabular}[c]{@{}c@{}}VolumeNet (D9, C32)\\ {[}ours{]}\end{tabular}   & \textbf{4.328} & \textbf{35.41} & \textbf{0.9845} & 0.483 & 0.230 \\
\hline \hline
\end{tabular}
\end{table}

\begin{table*}[t]
\renewcommand{\arraystretch}{1.5}
\caption{Quantitative results of lightweight models on IXI database. The compression ratios of parameters in each group are computed with their corresponding standard models. The result with underline is the best one in each group and the result in bold is the best one among all results. (The compression ratio of channels in CPD, LWv1, LWv2, and Queue module is set as 50\%, i.e., \(R=T/2)\)}
\label{table_4}
\centering
\begin{tabular}{cccccccc}
  \hline \hline
  Backbone Network & Lightweight   Method & RMSE & PSNR & SSIM & Time (s) & \#Param. (M) & \begin{tabular}[c]{@{}c@{}}Compression Ratio \\ of Parameters\end{tabular} \\
  \hline
  Tricubic & - & 6.752 & 31.54 & 0.9745 & - & - & - \\
  \hline
  \multirow{7}{*}{DCSRN \cite{chen2018brain}} & - & 4.789 & 34.53 & 0.9830 & 1.797 & 0.224 & 1 \\
    & Xception module & 4.809 & 34.49 & 0.9826 & 1.279 & 0.017 & 13 \\
    & MobileNetV2   module & \underline{4.663} & \underline{34.76} & \underline{0.9832} & 2.079 & 0.032 & 14 \\
    & CPD & 4.852 & 34.41 & 0.9824 & 1.151 & 0.012 & 19 \\
    & LWv1 & 4.861 & 34.40 & 0.9824 & 1.137 & 0.022 & 10 \\
    & LWv2 & 4.782 & 34.54 & 0.9828 & 0.970 & 0.023 & 9 \\
    & Queue module & 4.777 & 34.55 & 0.9828 & \underline{0.913} & 0.017 & 13 \\
  \hline
  \multirow{7}{*}{mDCSRN \cite{chen2018efficient}} & - & 4.667 & 34.75 & 0.9831 & 4.341 & 0.425 & 1 \\
    & Xception module & 4.655 & 34.77 & 0.9831 & 3.563 & 0.057 & 7 \\
    & MobileNetV2   module & 4.695 & 34.70 & 0.9830 & 3.837 & 0.076 & 6 \\
    & CPD & 4.721 & 34.65 & 0.9829 & 3.180 & 0.047 & 9 \\
    & LWv1 & 4.544 & 34.98 & 0.9836 & 3.180 & 0.064 & 7 \\
    & LWv2 & 4.518 & 35.03 & 0.9837 & 2.797 & 0.067 & 6 \\
    & Queue module & \underline{4.462} & \underline{35.14} & \underline{0.9839} & \underline{2.620} & 0.055 & 8 \\
    \hline
  \multirow{7}{*}{\begin{tabular}[c]{@{}c@{}}ParallelNet (D9, C32) \\ {[}ours{]}\end{tabular}} & - & 4.493 & 35.08 & 0.9838 & 1.335 & 2.566 & 1 \\
    & Xception module & 4.565 & 34.94 & 0.9836 & 2.312 & 0.178 & 14 \\
    & MobileNetV2   module & 4.541 & 34.99 & 0.9836 & 4.650 & 0.388 & 7 \\
    & CPD & 4.645 & 34.79 & 0.9832 & 1.736 & 0.133 & 19 \\
    & LWv1 & 4.384 & 35.29 & 0.9843 & 1.201 & 0.248 & 10 \\
    & LWv2 & 4.357 & 35.35 & 0.9844 & 0.486 & 0.280 & 9 \\
    & Queue module & \textbf{4.328} & \textbf{35.41} & \textbf{0.9845} & \textbf{0.483} & 0.230 & 11\\
    \hline \hline
  \end{tabular}
\end{table*}

TABLE III shows the quantitative results obtained by the proposed and state-of-the-art SR models on the IXI database. Here, we use root mean squared error (RMSE), PSNR, structural similarity index measure (SSIM) \cite{wang2004image}, prediction time, and parameters as quality evaluation indexes. Among the 3D SR models, the 3D SRCNN showed the smallest number of parameters; however, the accuracy of this model was not high due to its extremely shallow structure. DCSRN and mDCSRN, i.e., state-of-the-art CNN models for 3D medical volume data SR outperformed the 3D SRCNN. DCSRN demonstrated that a densely connected wide model can effectively improve accuracy. mDCSRN can be considered a complex model comprising multiple DCSRNs as basic modules. Although mDCSRN demonstrated improved accuracy compared to DCSRN, the number of network parameters and prediction time increased significantly. The proposed parallel network employs parallel connections to fuse hierarchical features, which improved the performance of the model. In addition, the construction method based on group convolution can expand the width of the model and avoid overfitting caused by too many parameters. In VolumeNet (D5, C32) and VolumeNet (D9, C32), the proposed Queue module was used to further optimize the ParallelNet structure. Although VolumeNet (D5, C32) gave the fastest prediction, excessive reduction in the number of parameters caused a slight decrease in accuracy. Therefore, we deepened the network, and then VolumeNet (D9, C32) outperformed all other models in terms of PSNR and SSIM, and the number of parameters was nearly the same as DCSRN. Although the proposed approaches are superior to the compare methods in terms of both PSNR and SSIM, PSNR and SSIM have different scales; thus, the improvement to SSIM is relatively small compared to that of PSNR.

Table IV shows that the proposed Queue module significantly improved the of 3D SR networks because the number of parameters were reduced by eight to 13 times. The lightweight module based on Xception used pointwise and depthwise convolutions successively without considering the channel information interaction, which results in reduced accuracy. The lightweight module based on MobileNetV2 used depthwise convolutions after expanding the channel, which maintained the accuracy of the shallow backbone (i.e., DCSRN). Nevertheless, the numbers of depthwise convolutions in MobileNetV2 required a long time to read the feature map, which resulted in slower processing in actual prediction. For comparison, we demonstrated that the use of CPD or lightweight structures based on depthwise convolutions (i.e., the CPD module and LWv1) in SR compressed a network to a great extent; however, this reduced accuracy. Note that we only changed the middle depthwise convolution layer in the LWv1 module to a cross-channel convolutional layer in the LWv2 module, and the accuracy and efficiency were improved. The convolution operation of Queue module was similar to that of LWv2; however, the independent pointwise layer could reduce the learning difficulty of cross-channel layers and deepen the network to improve accuracy. In addition, by comparing the CPD and Queue modules (or the LWv1 and LWv2 modules), we found that building a more efficient structure (i.e., changing depthwise layers to cross-channel convolutional layers) can effectively improve network accuracy and reduce the number of parameters simultaneously. Furthermore, the proposed lightweight module in VolumeNet is suitable for replacing standard 3D convolution layers; thus, it can be applied to any model to compress the CNN.

We visualize a comparison of performance, model size, and test time in Fig. 5. Here, to demonstrate different results more clearly, we attach magnified crops to the result images. As can be seen, VolumeNet achieved a good trade-off between accuracy, model size, and running time.

\begin{figure}[!t]
  \centering
  \includegraphics[width=3.4in]{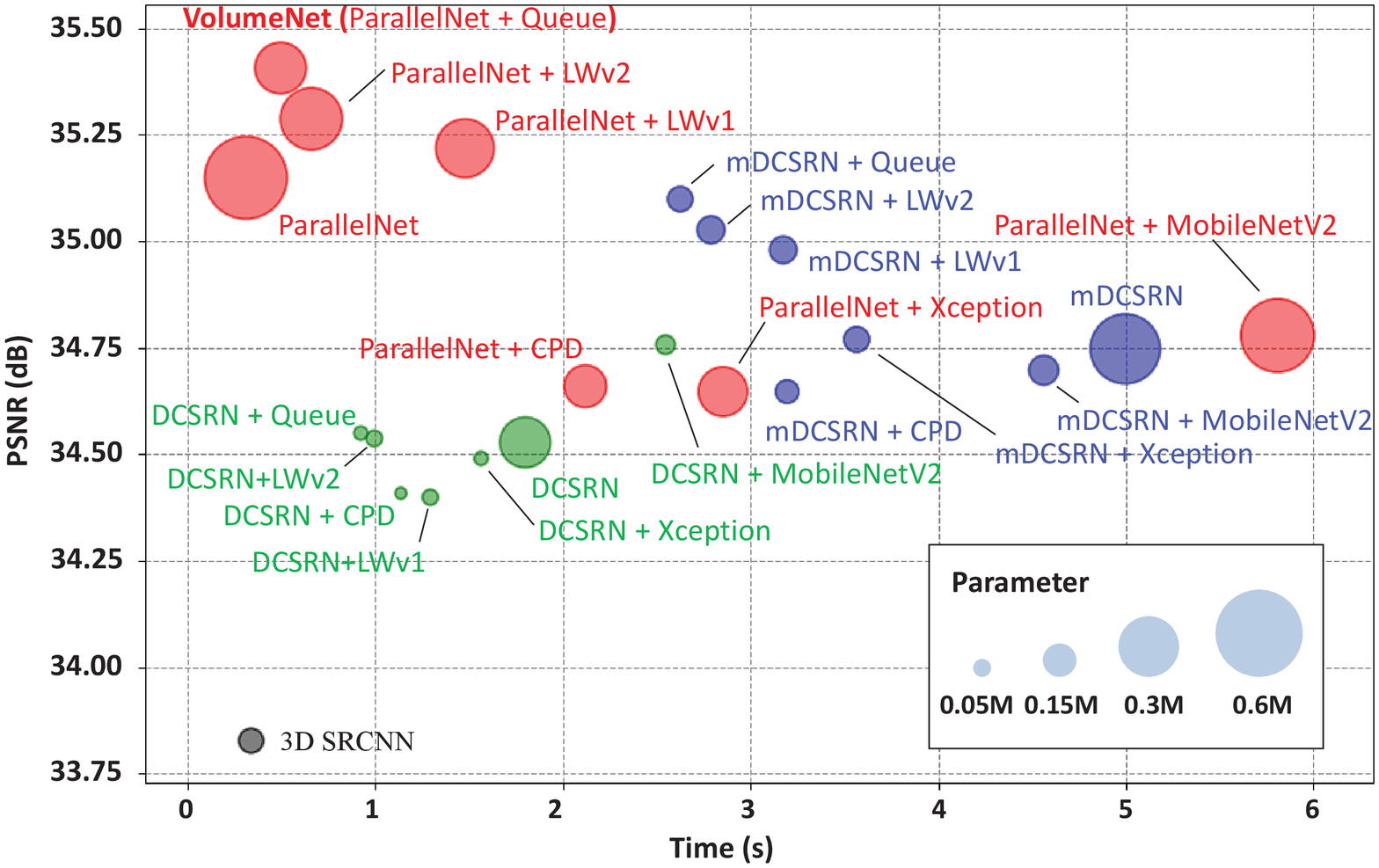}
  \caption{A comparison of the accuracy, processing speed, and model size on IXI database. The overall performance of the model near the upper left and with a smaller area is better.}
  \label{fig_5}
\end{figure}

Fig. 6 shows some examples of the SR results obtained by the proposed methods and state-of-the-art 3D CNNs on the test dataset. As can be seen, the SR images reconstructed by the proposed methods demonstrate the smallest difference from the GND images.

\begin{figure*}[!t]
  \centering
  \includegraphics[width=6in]{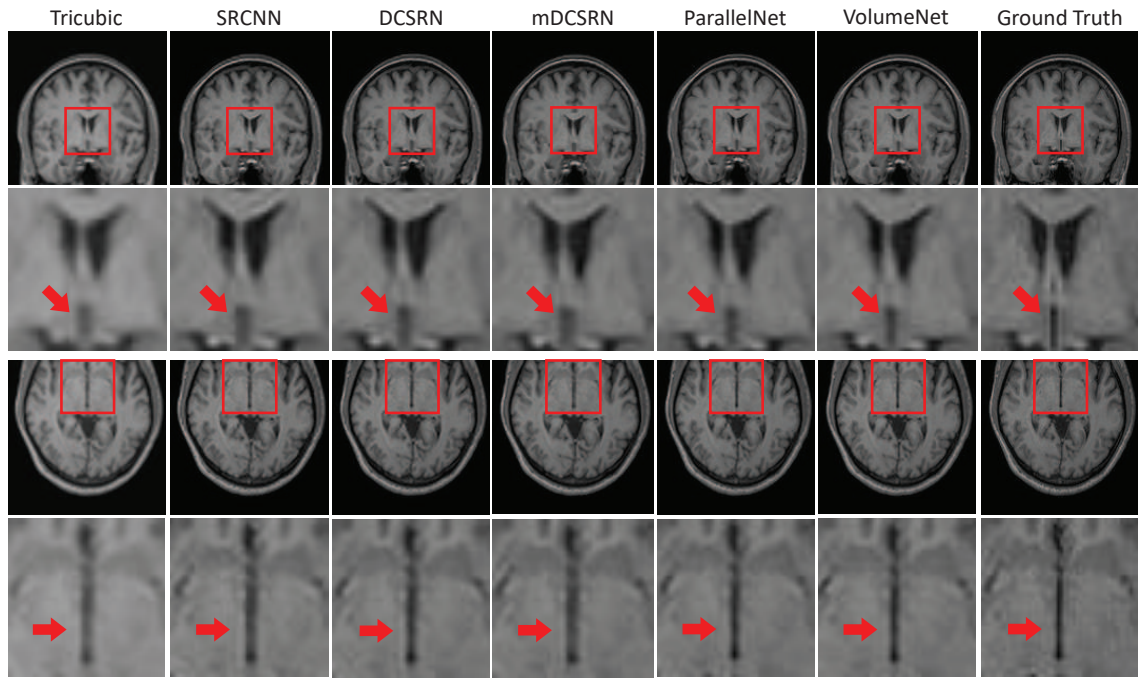}
  \caption{Illustration of SR results on IXI database with isotropic voxel upsampling (scale factor is ×2 in each direction and the compression ratio of the channels in VolumeNet is set as 50\%, i.e., \(R=T/2\)).}
  \label{fig_6}
\end{figure*}

\subsection{Comparison with Other Methods on LiTS database}

We used the training dataset of the Liver Tumor Segmentation (LiTS) database \cite{bilic2019liver} perform the CT image SR, which contains 130 CT scans. This database was acquired by different scanners and protocols from six different clinical sites with a largely varying in-plane resolution of 0.55 to 1.0 mm and slice spacing of 0.45 to 6.0 mm. Here, we randomly used 50 CT scans for training, 10 CT scans for validation, and 70 CT scans for testing. As preprocessing, we resized the data to \(250 \times 250 \times 150\) voxels and truncated all scans' image intensity values to the range [-1000, 1000] Hounsfield units to remove irrelevant details. We used these processed images as GND HR images and degraded them to LR images at \(\times 2\) scale in X, Y, and Z directions via tricubic interpolation.

For the training settings and loss function, here, we used the same settings as the MRI image SR (Section V‒A).

Fig. 7 and Table V show the qualitative and quantitative evaluation results of the proposed methods and several state-of-the-art 3D CNNs on the LiTS database. As shown in Fig. 7, the proposed VolumeNet performed the best, followed by ParallelNet. Compared to the existing methods, the proposed method fully integrated deep features. Thus, the proposed methods reconstructed sharper and more accurate results. In addition, compared to the IXI database, the LiTS database has a more uneven distribution of data brightness, making it comparatively more difficult to restore. Our experimental results have proven the superiority of the proposed method relative to dealing with images with different modalities and unbalanced pixel value distribution.

\begin{table}[!t]
  \renewcommand{\arraystretch}{1.5}
  \caption{Quantitative results of CT images on LiTS database.}
  \label{table_5}
  \centering
  \begin{tabular}{cccccc}
  \hline \hline
  Method & RMSE &	PSNR & SSIM & \begin{tabular}[c]{@{}c@{}}Time\\ (s)\end{tabular} & \begin{tabular}[c]{@{}c@{}}\#param.\\ (M)\end{tabular}\\
  \hline
  Tricubic interpolation & 6.998 & 31.23 & 0.9255 &	- &	-\\
  \hline
  3D SRCNN \cite{pham2017brain} & 5.672 & 33.06 & 0.9302 &	0.305 &	0.053\\
  \hline
  DCSRN \cite{chen2018brain} & 3.914 &	36.28 &	0.9536 & 1.564 & 0.224\\
  \hline
  mDCSRN \cite{chen2018efficient} & 3.816 &	36.50 &	0.9568 & 4.335 & 0.425\\
  \hline
  \begin{tabular}[c]{@{}c@{}}ParallelNet (D5, C32) \\ {[}ours{]}\end{tabular} & 3.637 &	36.92 &	0.9594 & 0.304 & 0.595\\
  \begin{tabular}[c]{@{}c@{}}VolumeNet (D5, C32)\\ {[}ours{]}\end{tabular}   & 3.839 & 36.45 & 0.9571 & \textbf{0.208} & 0.089 \\
  \begin{tabular}[c]{@{}c@{}}VolumeNet (D9, C32)\\ {[}ours{]}\end{tabular}   & \textbf{3.569} & \textbf{37.08} & \textbf{0.9601} & 0.493 & 0.230 \\
  \hline \hline
  \end{tabular}
\end{table}

\begin{figure*}[!t]
  \centering
  \includegraphics[width=6in]{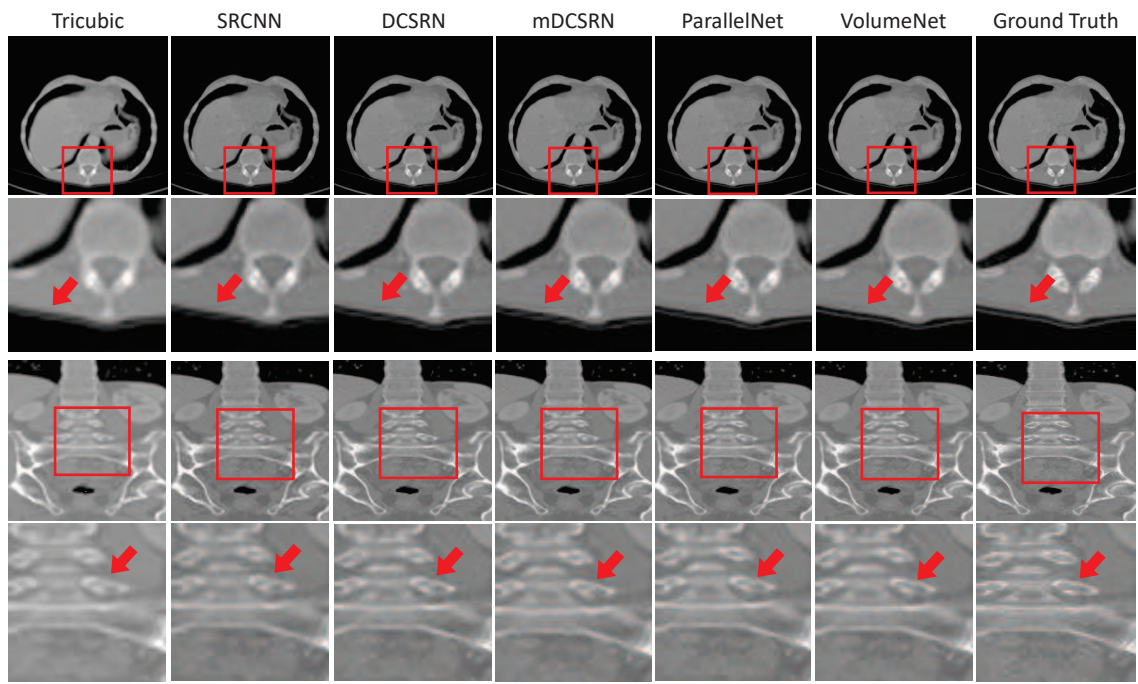}
  \caption{Illustration of SR results on LiTS database with isotropic voxel upsampling (scale factor is ×2 in each direction and the compression ratio of the channels in VolumeNet is set as 50\%, i.e., \(R=T/2\)}
  \label{fig_7}
\end{figure*}

\section{Conclusion}

In this paper, we have proposed a 3D CNN model for medical volumetric data SR named ParallelNet and an efficient ParallelNet called VolumeNet. We have also demonstrated the lightweight Queue module's superiority when combining it with several typical state-of-the-art SR models. The proposed Queue module can reduce the network parameters significantly; thus, the computational cost is also reduced. In addition, we have demonstrated that the proposed VolumeNet outperformed state-of-the-art methods in terms of both accuracy and efficiency; thus, it is suitable for practical application, e.g., such as small training samples and short prediction time.

\section*{Acknowledgment}

This work is supported in part by Japan Society for Promotion of Science (JSPS) under Grant No. 19J13820, the Grant-in-Aid for Young Scientists (18K18078), Grant-in-Aid for Scientific Research (B) (18H03267) and Grant-in-Aid for Challenging Research (Exploratory) (20K21821).



%





\ifCLASSOPTIONcaptionsoff
  \newpage
\fi



\bibliographystyle{IEEEtran}
\bibliography{reference}

\begin{thebibliography}{10}
\providecommand{\url}[1]{#1}
\csname url@samestyle\endcsname
\providecommand{\newblock}{\relax}
\providecommand{\bibinfo}[2]{#2}
\providecommand{\BIBentrySTDinterwordspacing}{\spaceskip=0pt\relax}
\providecommand{\BIBentryALTinterwordstretchfactor}{4}
\providecommand{\BIBentryALTinterwordspacing}{\spaceskip=\fontdimen2\font plus
\BIBentryALTinterwordstretchfactor\fontdimen3\font minus
  \fontdimen4\font\relax}
\providecommand{\BIBforeignlanguage}[2]{{%
\expandafter\ifx\csname l@#1\endcsname\relax
\typeout{** WARNING: IEEEtran.bst: No hyphenation pattern has been}%
\typeout{** loaded for the language `#1'. Using the pattern for}%
\typeout{** the default language instead.}%
\else
\language=\csname l@#1\endcsname
\fi
#2}}
\providecommand{\BIBdecl}{\relax}
\BIBdecl

\bibitem{ozyurt2020expert}
F.~{\"O}zyurt, E.~Sert, and D.~Avc{\i}, ``An expert system for brain tumor
  detection: Fuzzy c-means with super resolution and convolutional neural
  network with extreme learning machine,'' \emph{Medical hypotheses}, vol. 134,
  p. 109433, 2020.

\bibitem{hatvani2018tensor}
J.~Hatvani, A.~Basarab, J.-Y. Tourneret, M.~Gy{\"o}ngy, and D.~Kouam{\'e}, ``A
  tensor factorization method for 3-d super resolution with application to
  dental ct,'' \emph{IEEE transactions on medical imaging}, vol.~38, no.~6, pp.
  1524--1531, 2018.

\bibitem{mahapatra2019image}
D.~Mahapatra, B.~Bozorgtabar, and R.~Garnavi, ``Image super-resolution using
  progressive generative adversarial networks for medical image analysis,''
  \emph{Computerized Medical Imaging and Graphics}, vol.~71, pp. 30--39, 2019.

\bibitem{anoop2020super}
V.~Anoop and P.~Bipin, ``Super-resolution based automatic diagnosis of retinal
  disease detection for clinical applications,'' \emph{Neural Processing
  Letters}, vol.~52, no.~2, pp. 1155--1170, 2020.

\bibitem{pham2017brain}
C.-H. Pham, A.~Ducournau, R.~Fablet, and F.~Rousseau, ``Brain mri
  super-resolution using deep 3d convolutional networks,'' in \emph{2017 IEEE
  14th International Symposium on Biomedical Imaging (ISBI 2017)}.\hskip 1em
  plus 0.5em minus 0.4em\relax IEEE, 2017, pp. 197--200.

\bibitem{chen2018brain}
Y.~Chen, Y.~Xie, Z.~Zhou, F.~Shi, A.~G. Christodoulou, and D.~Li, ``Brain mri
  super resolution using 3d deep densely connected neural networks,'' in
  \emph{2018 IEEE 15th International Symposium on Biomedical Imaging (ISBI
  2018)}.\hskip 1em plus 0.5em minus 0.4em\relax IEEE, 2018, pp. 739--742.

\bibitem{chen2018efficient}
Y.~Chen, F.~Shi, A.~G. Christodoulou, Y.~Xie, Z.~Zhou, and D.~Li, ``Efficient
  and accurate mri super-resolution using a generative adversarial network and
  3d multi-level densely connected network,'' in \emph{International Conference
  on Medical Image Computing and Computer-Assisted Intervention}.\hskip 1em
  plus 0.5em minus 0.4em\relax Springer, 2018, pp. 91--99.

\bibitem{zhu2019can}
J.~Zhu, G.~Yang, and P.~Lio, ``How can we make gan perform better in single
  medical image super-resolution? a lesion focused multi-scale approach,'' in
  \emph{2019 IEEE 16th International Symposium on Biomedical Imaging (ISBI
  2019)}.\hskip 1em plus 0.5em minus 0.4em\relax IEEE, 2019, pp. 1669--1673.

\bibitem{zhao2019channel}
X.~Zhao, Y.~Zhang, T.~Zhang, and X.~Zou, ``Channel splitting network for single
  mr image super-resolution,'' \emph{IEEE Transactions on Image Processing},
  vol.~28, no.~11, pp. 5649--5662, 2019.

\bibitem{you2019ct}
C.~You, G.~Li, Y.~Zhang, X.~Zhang, H.~Shan, M.~Li, S.~Ju, Z.~Zhao, Z.~Zhang,
  W.~Cong \emph{et~al.}, ``Ct super-resolution gan constrained by the
  identical, residual, and cycle learning ensemble (gan-circle),'' \emph{IEEE
  transactions on medical imaging}, vol.~39, no.~1, pp. 188--203, 2019.

\bibitem{du2020super}
J.~Du, Z.~He, L.~Wang, A.~Gholipour, Z.~Zhou, D.~Chen, and Y.~Jia,
  ``Super-resolution reconstruction of single anisotropic 3d mr images using
  residual convolutional neural network,'' \emph{Neurocomputing}, vol. 392, pp.
  209--220, 2020.

\bibitem{lyu2020mri}
Q.~Lyu, H.~Shan, and G.~Wang, ``Mri super-resolution with ensemble learning and
  complementary priors,'' \emph{IEEE Transactions on Computational Imaging},
  vol.~6, pp. 615--624, 2020.

\bibitem{du2018accelerated}
J.~Du, L.~Wang, A.~Gholipour, Z.~He, and Y.~Jia, ``Accelerated super-resolution
  mr image reconstruction via a 3d densely connected deep convolutional neural
  network,'' in \emph{2018 IEEE International Conference on Bioinformatics and
  Biomedicine (BIBM)}.\hskip 1em plus 0.5em minus 0.4em\relax IEEE, 2018, pp.
  349--355.

\bibitem{dong2015image}
C.~Dong, C.~C. Loy, K.~He, and X.~Tang, ``Image super-resolution using deep
  convolutional networks,'' \emph{IEEE transactions on pattern analysis and
  machine intelligence}, vol.~38, no.~2, pp. 295--307, 2015.

\bibitem{kim2016accurate}
J.~Kim, J.~Kwon~Lee, and K.~Mu~Lee, ``Accurate image super-resolution using
  very deep convolutional networks,'' in \emph{Proceedings of the IEEE
  conference on computer vision and pattern recognition}, 2016, pp. 1646--1654.

\bibitem{shi2016real}
W.~Shi, J.~Caballero, F.~Husz{\'a}r, J.~Totz, A.~P. Aitken, R.~Bishop,
  D.~Rueckert, and Z.~Wang, ``Real-time single image and video super-resolution
  using an efficient sub-pixel convolutional neural network,'' in
  \emph{Proceedings of the IEEE conference on computer vision and pattern
  recognition}, 2016, pp. 1874--1883.

\bibitem{zhang2018residual}
Y.~Zhang, Y.~Tian, Y.~Kong, B.~Zhong, and Y.~Fu, ``Residual dense network for
  image super-resolution,'' in \emph{Proceedings of the IEEE conference on
  computer vision and pattern recognition}, 2018, pp. 2472--2481.

\bibitem{zhang2018image}
Y.~Zhang, K.~Li, K.~Li, L.~Wang, B.~Zhong, and Y.~Fu, ``Image super-resolution
  using very deep residual channel attention networks,'' in \emph{Proceedings
  of the European Conference on Computer Vision (ECCV)}, 2018, pp. 286--301.

\bibitem{mao2016image}
X.~Mao, C.~Shen, and Y.-B. Yang, ``Image restoration using very deep
  convolutional encoder-decoder networks with symmetric skip connections,'' in
  \emph{Advances in neural information processing systems}, 2016, pp.
  2802--2810.

\bibitem{han2018image}
W.~Han, S.~Chang, D.~Liu, M.~Yu, M.~Witbrock, and T.~S. Huang, ``Image
  super-resolution via dual-state recurrent networks,'' in \emph{Proceedings of
  the IEEE conference on computer vision and pattern recognition}, 2018, pp.
  1654--1663.

\bibitem{li2018multi}
J.~Li, F.~Fang, K.~Mei, and G.~Zhang, ``Multi-scale residual network for image
  super-resolution,'' in \emph{Proceedings of the European Conference on
  Computer Vision (ECCV)}, 2018, pp. 517--532.

\bibitem{huang2017densely}
G.~Huang, Z.~Liu, L.~Van Der~Maaten, and K.~Q. Weinberger, ``Densely connected
  convolutional networks,'' in \emph{Proceedings of the IEEE conference on
  computer vision and pattern recognition}, 2017, pp. 4700--4708.

\bibitem{ledig2017photo}
C.~Ledig, L.~Theis, F.~Husz{\'a}r, J.~Caballero, A.~Cunningham, A.~Acosta,
  A.~Aitken, A.~Tejani, J.~Totz, Z.~Wang \emph{et~al.}, ``Photo-realistic
  single image super-resolution using a generative adversarial network,'' in
  \emph{Proceedings of the IEEE conference on computer vision and pattern
  recognition}, 2017, pp. 4681--4690.

\bibitem{zhang2018learning}
K.~Zhang, W.~Zuo, and L.~Zhang, ``Learning a single convolutional
  super-resolution network for multiple degradations,'' in \emph{Proceedings of
  the IEEE Conference on Computer Vision and Pattern Recognition}, 2018, pp.
  3262--3271.

\bibitem{haris2018deep}
M.~Haris, G.~Shakhnarovich, and N.~Ukita, ``Deep back-projection networks for
  super-resolution,'' in \emph{Proceedings of the IEEE conference on computer
  vision and pattern recognition}, 2018, pp. 1664--1673.

\bibitem{tong2017image}
T.~Tong, G.~Li, X.~Liu, and Q.~Gao, ``Image super-resolution using dense skip
  connections,'' in \emph{Proceedings of the IEEE International Conference on
  Computer Vision}, 2017, pp. 4799--4807.

\bibitem{kim2016deeply}
J.~Kim, J.~Kwon~Lee, and K.~Mu~Lee, ``Deeply-recursive convolutional network
  for image super-resolution,'' in \emph{Proceedings of the IEEE conference on
  computer vision and pattern recognition}, 2016, pp. 1637--1645.

\bibitem{lai2018fast}
W.-S. Lai, J.-B. Huang, N.~Ahuja, and M.-H. Yang, ``Fast and accurate image
  super-resolution with deep laplacian pyramid networks,'' \emph{IEEE
  transactions on pattern analysis and machine intelligence}, vol.~41, no.~11,
  pp. 2599--2613, 2018.

\bibitem{lai2017deep}
------, ``Deep laplacian pyramid networks for fast and accurate
  super-resolution,'' in \emph{Proceedings of the IEEE conference on computer
  vision and pattern recognition}, 2017, pp. 624--632.

\bibitem{dahl2017pixel}
R.~Dahl, M.~Norouzi, and J.~Shlens, ``Pixel recursive super resolution,'' in
  \emph{Proceedings of the IEEE international conference on computer vision},
  2017, pp. 5439--5448.

\bibitem{ren2017image}
H.~Ren, M.~El-Khamy, and J.~Lee, ``Image super resolution based on fusing
  multiple convolution neural networks,'' in \emph{Proceedings of the IEEE
  Conference on Computer Vision and Pattern Recognition Workshops}, 2017, pp.
  54--61.

\bibitem{wang2018fully}
Y.~Wang, F.~Perazzi, B.~McWilliams, A.~Sorkine-Hornung, O.~Sorkine-Hornung, and
  C.~Schroers, ``A fully progressive approach to single-image
  super-resolution,'' in \emph{Proceedings of the IEEE Conference on Computer
  Vision and Pattern Recognition Workshops}, 2018, pp. 864--873.

\bibitem{bae2017beyond}
W.~Bae, J.~Yoo, and J.~Chul~Ye, ``Beyond deep residual learning for image
  restoration: Persistent homology-guided manifold simplification,'' in
  \emph{Proceedings of the IEEE conference on computer vision and pattern
  recognition workshops}, 2017, pp. 145--153.

\bibitem{guo2017deep}
T.~Guo, H.~Seyed~Mousavi, T.~Huu~Vu, and V.~Monga, ``Deep wavelet prediction
  for image super-resolution,'' in \emph{Proceedings of the IEEE Conference on
  Computer Vision and Pattern Recognition Workshops}, 2017, pp. 104--113.

\bibitem{huang2017wavelet}
H.~Huang, R.~He, Z.~Sun, and T.~Tan, ``Wavelet-srnet: A wavelet-based cnn for
  multi-scale face super resolution,'' in \emph{Proceedings of the IEEE
  International Conference on Computer Vision}, 2017, pp. 1689--1697.

\bibitem{liu2018multi}
P.~Liu, H.~Zhang, K.~Zhang, L.~Lin, and W.~Zuo, ``Multi-level wavelet-cnn for
  image restoration,'' in \emph{Proceedings of the IEEE conference on computer
  vision and pattern recognition workshops}, 2018, pp. 773--782.

\bibitem{wang2020deep}
Z.~Wang, J.~Chen, and S.~C. Hoi, ``Deep learning for image super-resolution: A
  survey,'' \emph{IEEE Transactions on Pattern Analysis and Machine
  Intelligence}, 2020.

\bibitem{tan2019efficientnet}
M.~Tan and Q.~Le, ``Efficientnet: Rethinking model scaling for convolutional
  neural networks,'' in \emph{International Conference on Machine Learning},
  2019, pp. 6105--6114.

\bibitem{zhao2018self}
C.~Zhao, A.~Carass, B.~E. Dewey, and J.~L. Prince, ``Self super-resolution for
  magnetic resonance images using deep networks,'' in \emph{2018 IEEE 15th
  International Symposium on Biomedical Imaging (ISBI 2018)}.\hskip 1em plus
  0.5em minus 0.4em\relax IEEE, 2018, pp. 365--368.

\bibitem{shi2018mr}
J.~Shi, Z.~Li, S.~Ying, C.~Wang, Q.~Liu, Q.~Zhang, and P.~Yan, ``Mr image
  super-resolution via wide residual networks with fixed skip connection,''
  \emph{IEEE journal of biomedical and health informatics}, vol.~23, no.~3, pp.
  1129--1140, 2018.

\bibitem{zhao2016deep}
L.~Zhao, J.~Wang, X.~Li, Z.~Tu, and W.~Zeng, ``Deep convolutional neural
  networks with merge-and-run mappings,'' \emph{arXiv preprint
  arXiv:1611.07718}, 2016.

\bibitem{hu2018single}
Y.~Hu, X.~Gao, J.~Li, Y.~Huang, and H.~Wang, ``Single image super-resolution
  via cascaded multi-scale cross network,'' \emph{arXiv preprint
  arXiv:1802.08808}, 2018.

\bibitem{cherukuri2019deep}
V.~Cherukuri, T.~Guo, S.~J. Schiff, and V.~Monga, ``Deep mr brain image
  super-resolution using spatio-structural priors,'' \emph{IEEE Transactions on
  Image Processing}, vol.~29, pp. 1368--1383, 2019.

\bibitem{rigamonti2013learning}
R.~Rigamonti, A.~Sironi, V.~Lepetit, and P.~Fua, ``Learning separable
  filters,'' in \emph{Proceedings of the IEEE conference on computer vision and
  pattern recognition}, 2013, pp. 2754--2761.

\bibitem{kim2015compression}
Y.-D. Kim, E.~Park, S.~Yoo, T.~Choi, L.~Yang, and D.~Shin, ``Compression of
  deep convolutional neural networks for fast and low power mobile
  applications,'' \emph{arXiv preprint arXiv:1511.06530}, 2015.

\bibitem{lebedev2014speeding}
V.~Lebedev, Y.~Ganin, M.~Rakhuba, I.~Oseledets, and V.~Lempitsky, ``Speeding-up
  convolutional neural networks using fine-tuned cp-decomposition,''
  \emph{arXiv preprint arXiv:1412.6553}, 2014.

\bibitem{howard2017mobilenets}
A.~G. Howard, M.~Zhu, B.~Chen, D.~Kalenichenko, W.~Wang, T.~Weyand,
  M.~Andreetto, and H.~Adam, ``Mobilenets: Efficient convolutional neural
  networks for mobile vision applications,'' \emph{arXiv preprint
  arXiv:1704.04861}, 2017.

\bibitem{chollet2017xception}
F.~Chollet, ``Xception: Deep learning with depthwise separable convolutions,''
  in \emph{Proceedings of the IEEE conference on computer vision and pattern
  recognition}, 2017, pp. 1251--1258.

\bibitem{wang2017factorized}
M.~Wang, B.~Liu, and H.~Foroosh, ``Factorized convolutional neural networks,''
  in \emph{Proceedings of the IEEE International Conference on Computer Vision
  Workshops}, 2017, pp. 545--553.

\bibitem{jin2014flattened}
J.~Jin, A.~Dundar, and E.~Culurciello, ``Flattened convolutional neural
  networks for feedforward acceleration,'' \emph{arXiv preprint
  arXiv:1412.5474}, 2014.

\bibitem{tai2017memnet}
Y.~Tai, J.~Yang, X.~Liu, and C.~Xu, ``Memnet: A persistent memory network for
  image restoration,'' in \emph{Proceedings of the IEEE international
  conference on computer vision}, 2017, pp. 4539--4547.

\bibitem{tai2017image}
Y.~Tai, J.~Yang, and X.~Liu, ``Image super-resolution via deep recursive
  residual network,'' in \emph{Proceedings of the IEEE conference on computer
  vision and pattern recognition}, 2017, pp. 3147--3155.

\bibitem{hui2018fast}
Z.~Hui, X.~Wang, and X.~Gao, ``Fast and accurate single image super-resolution
  via information distillation network,'' in \emph{Proceedings of the IEEE
  conference on computer vision and pattern recognition}, 2018, pp. 723--731.

\bibitem{ahn2018fast}
N.~Ahn, B.~Kang, and K.-A. Sohn, ``Fast, accurate, and lightweight
  super-resolution with cascading residual network,'' in \emph{Proceedings of
  the European Conference on Computer Vision (ECCV)}, 2018, pp. 252--268.

\bibitem{sandler2018mobilenetv2}
M.~Sandler, A.~Howard, M.~Zhu, A.~Zhmoginov, and L.-C. Chen, ``Mobilenetv2:
  Inverted residuals and linear bottlenecks,'' in \emph{Proceedings of the IEEE
  conference on computer vision and pattern recognition}, 2018, pp. 4510--4520.

\bibitem{tran2018closer}
D.~Tran, H.~Wang, L.~Torresani, J.~Ray, Y.~LeCun, and M.~Paluri, ``A closer
  look at spatiotemporal convolutions for action recognition,'' in
  \emph{Proceedings of the IEEE conference on Computer Vision and Pattern
  Recognition}, 2018, pp. 6450--6459.

\bibitem{awasthi2020deep}
N.~Awasthi, G.~Jain, S.~K. Kalva, M.~Pramanik, and P.~K. Yalavarthy, ``Deep
  neural network based sinogram super-resolution and bandwidth enhancement for
  limited-data photoacoustic tomography,'' \emph{IEEE Transactions on
  Ultrasonics, Ferroelectrics, and Frequency Control}, 2020.

\bibitem{he2016deep}
K.~He, X.~Zhang, S.~Ren, and J.~Sun, ``Deep residual learning for image
  recognition,'' in \emph{Proceedings of the IEEE conference on computer vision
  and pattern recognition}, 2016, pp. 770--778.

\bibitem{lebedev2018speeding}
V.~Lebedev and V.~Lempitsky, ``Speeding-up convolutional neural networks: A
  survey,'' \emph{Bulletin of the Polish Academy of Sciences. Technical
  Sciences}, vol.~66, no.~6, 2018.

\bibitem{szegedy2015going}
C.~Szegedy, W.~Liu, Y.~Jia, P.~Sermanet, S.~Reed, D.~Anguelov, D.~Erhan,
  V.~Vanhoucke, and A.~Rabinovich, ``Going deeper with convolutions,'' in
  \emph{Proceedings of the IEEE conference on computer vision and pattern
  recognition}, 2015, pp. 1--9.

\bibitem{szegedy2016inception}
C.~Szegedy, S.~Ioffe, V.~Vanhoucke, and A.~Alemi, ``Inception-v4,
  inception-resnet and the impact of residual connections on learning,''
  \emph{arXiv preprint arXiv:1602.07261}, 2016.

\bibitem{zhang2018shufflenet}
X.~Zhang, X.~Zhou, M.~Lin, and J.~Sun, ``Shufflenet: An extremely efficient
  convolutional neural network for mobile devices,'' in \emph{Proceedings of
  the IEEE conference on computer vision and pattern recognition}, 2018, pp.
  6848--6856.

\bibitem{freeman2018effnet}
I.~Freeman, L.~Roese-Koerner, and A.~Kummert, ``Effnet: An efficient structure
  for convolutional neural networks,'' in \emph{2018 25th IEEE International
  Conference on Image Processing (ICIP)}.\hskip 1em plus 0.5em minus
  0.4em\relax IEEE, 2018, pp. 6--10.

\bibitem{wu2018shift}
B.~Wu, A.~Wan, X.~Yue, P.~Jin, S.~Zhao, N.~Golmant, A.~Gholaminejad,
  J.~Gonzalez, and K.~Keutzer, ``Shift: A zero flop, zero parameter alternative
  to spatial convolutions,'' in \emph{Proceedings of the IEEE Conference on
  Computer Vision and Pattern Recognition}, 2018, pp. 9127--9135.

\bibitem{lin2013network}
M.~Lin, Q.~Chen, and S.~Yan, ``Network in network,'' \emph{arXiv preprint
  arXiv:1312.4400}, 2013.

\bibitem{iandola2016squeezenet}
F.~N. Iandola, S.~Han, M.~W. Moskewicz, K.~Ashraf, W.~J. Dally, and K.~Keutzer,
  ``Squeezenet: Alexnet-level accuracy with 50x fewer parameters and< 0.5 mb
  model size,'' \emph{arXiv preprint arXiv:1602.07360}, 2016.

\bibitem{krizhevsky2012imagenet}
A.~Krizhevsky, I.~Sutskever, and G.~E. Hinton, ``Imagenet classification with
  deep convolutional neural networks,'' in \emph{Advances in neural information
  processing systems}, 2012, pp. 1097--1105.

\bibitem{xie2017aggregated}
S.~Xie, R.~Girshick, P.~Doll{\'a}r, Z.~Tu, and K.~He, ``Aggregated residual
  transformations for deep neural networks,'' in \emph{Proceedings of the IEEE
  conference on computer vision and pattern recognition}, 2017, pp. 1492--1500.

\bibitem{glorot2011deep}
X.~Glorot, A.~Bordes, and Y.~Bengio, ``Deep sparse rectifier neural networks,''
  in \emph{Proceedings of the fourteenth international conference on artificial
  intelligence and statistics}, 2011, pp. 315--323.

\bibitem{lim2017enhanced}
B.~Lim, S.~Son, H.~Kim, S.~Nah, and K.~Mu~Lee, ``Enhanced deep residual
  networks for single image super-resolution,'' in \emph{Proceedings of the
  IEEE conference on computer vision and pattern recognition workshops}, 2017,
  pp. 136--144.

\bibitem{zhao2016loss}
H.~Zhao, O.~Gallo, I.~Frosio, and J.~Kautz, ``Loss functions for image
  restoration with neural networks,'' \emph{IEEE Transactions on computational
  imaging}, vol.~3, no.~1, pp. 47--57, 2016.

\bibitem{ixi}
\url{https://brain-development.org/ixi-dataset/}.

\bibitem{kingma2014adam}
D.~P. Kingma and J.~Ba, ``Adam: A method for stochastic optimization,''
  \emph{arXiv preprint arXiv:1412.6980}, 2014.

\bibitem{wang2004image}
Z.~Wang, A.~C. Bovik, H.~R. Sheikh, and E.~P. Simoncelli, ``Image quality
  assessment: from error visibility to structural similarity,'' \emph{IEEE
  transactions on image processing}, vol.~13, no.~4, pp. 600--612, 2004.

\bibitem{bilic2019liver}
P.~Bilic, P.~F. Christ, E.~Vorontsov, G.~Chlebus, H.~Chen, Q.~Dou, C.-W. Fu,
  X.~Han, P.-A. Heng, J.~Hesser \emph{et~al.}, ``The liver tumor segmentation
  benchmark (lits),'' \emph{arXiv preprint arXiv:1901.04056}, 2019.

\end{thebibliography}
\end{document}